\documentclass{elsart5p}

\usepackage{algorithm}
\usepackage{algpseudocode}
\usepackage{amsmath}
\usepackage{amssymb}
\usepackage{graphicx}

\usepackage{caption}
\usepackage{subcaption}

\newcommand{\eat}[1]{}

\begin{document}

\begin{frontmatter}



\title{Incorporating Sharp Features in the General Solid Sweep Framework}

\author{Bharat Adsul, Jinesh Machchhar, Milind Sohoni}

\begin{abstract}
This paper extends a recently proposed robust computational
 framework for constructing
 the boundary representation (brep) of the volume
 swept by a given smooth solid moving along a one parameter
 family $h$ of rigid motions.  Our extension allows the input solid
 to have sharp features, i.e., to be of class G0 wherein, the 
 unit outward normal to the solid may be
 discontinuous. In the earlier framework, the solid to be swept was
 restricted to be G1, and
 thus this is a significant and useful extension of that work. 

 This naturally requires a precise description of the geometry of 
 the surface generated by the sweep of a sharp edge supported by two 
 intersecting smooth faces. We uncover the geometry along with the related issues like
 parametrization, self-intersection and singularities via a novel mathematical analysis.
 Correct trimming of such a surface is achieved by a delicate analysis of the interplay between the 
 cone of normals at a sharp point and its trajectory under $h$. The overall topology
 is explicated by a key lifting theorem which allows us to
 compute the adjacency relations amongst entities in the swept volume by
 relating them to corresponding adjacencies in the input solid. 
 Moreover, global issues related to body-check such as orientation are efficiently resolved.  
 Many examples from a pilot implementation illustrate the efficiency and
 effectiveness of our framework.

\end{abstract}

\begin{keyword}
Solid sweep, swept volume, boundary representation, solid modeling, G0-solids, parametric curves and surfaces
\end{keyword}
\end{frontmatter}

\section{Introduction} 	\label{introSec}

\begin{figure}
 \centering
 \includegraphics[scale=0.3]{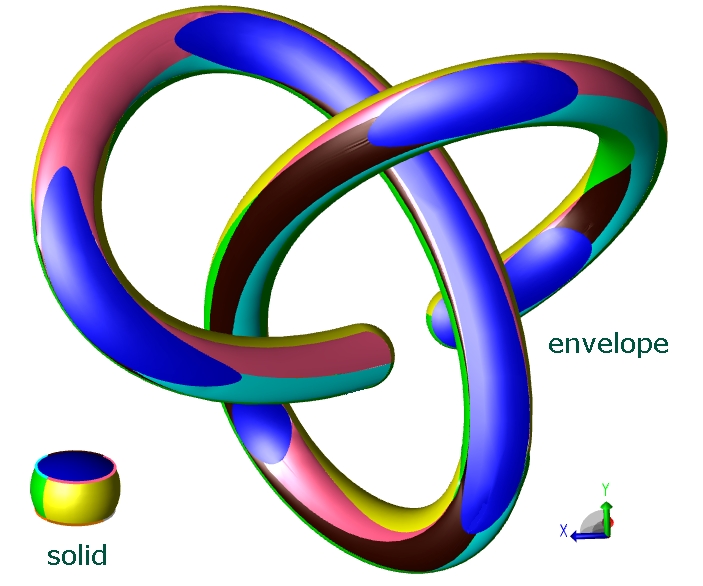}
\caption{An example of swept volume.}
 \label{trefoilFig}
\end{figure}

\begin{figure}
 \centering
 \includegraphics[scale=0.4]{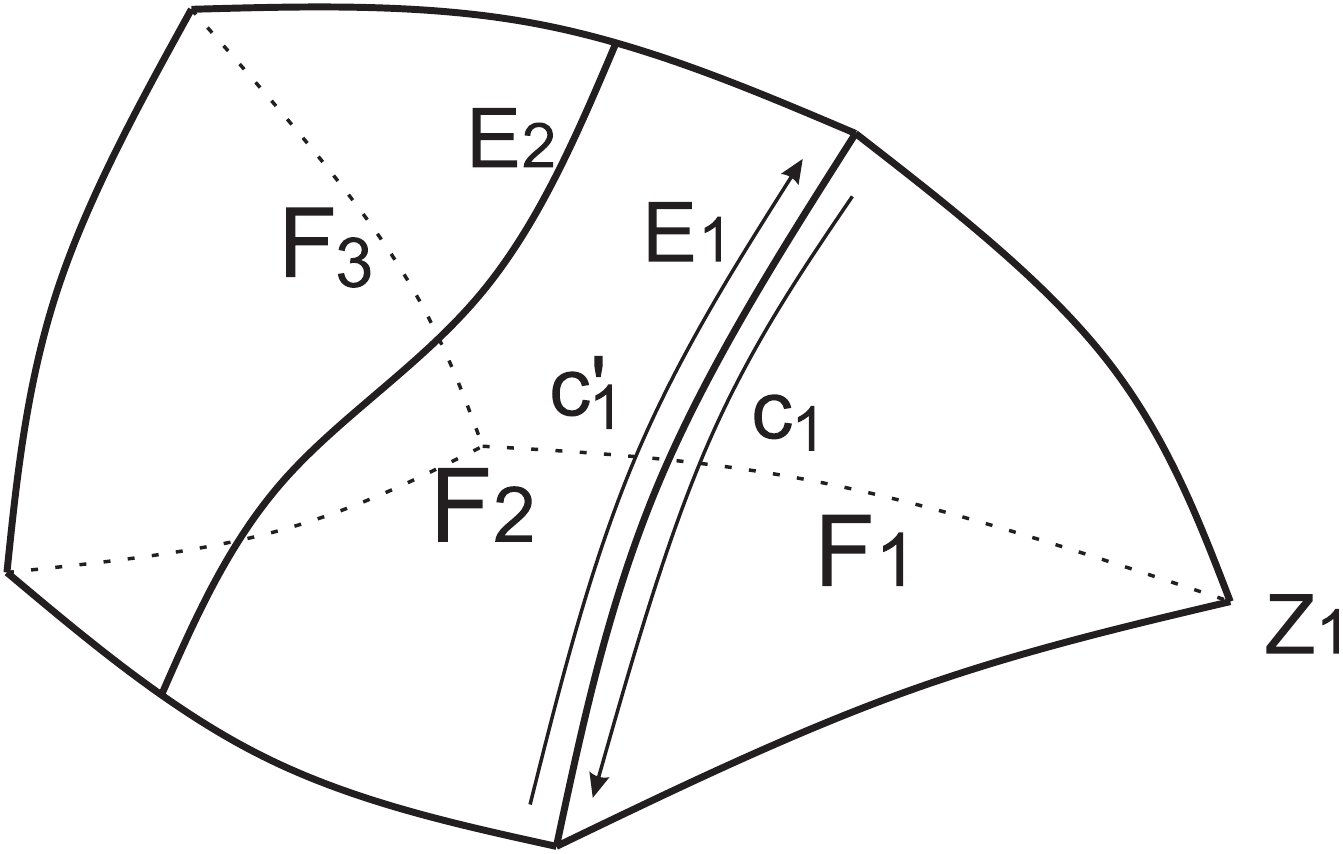}
 \caption{Brep of a solid.}
 \label{solidFig}
\end{figure}

In this paper we investigate the computation of the volume swept by a given solid moving along a smooth one parameter 
family of rigid motions.  We assume the solid to be of class G0, wherein, the unit outward normal may be discontinuous 
at the intersection of two or more faces.
Solid sweep has several applications, viz. CNC-machining verification~\cite{completeSweep, completeSweep2}, 
collision detection, motion planning~\cite{survey} and packaging~\cite{scroll}.  An example of solid sweep appears in Fig.~\ref{trefoilFig}.
We adopt the industry standard parametric boundary representation (brep) format to input the solid 
and output the swept volume.  In the brep format, the solid $M$ is represented by its boundary $\partial M$ which separates the 
interior of $M$ from its exterior.  The brep of $M$ consists of the parametric definitions of the faces, edges and vertices as well 
as their orientations and adjacency relations amongst these.  Fig.~\ref{solidFig} schematically illustrates such a solid.  

The computation of swept volume has been extensively studied~\cite{jacobian, sede, errorBounds, classifyPoints, dual, planarSweep, polyhedra, 3dSweep}.
 In~\cite{jacobian},  
the envelope ${\cal E}$ is modeled as the the set of points where the Jacobian of the sweep map is rank deficient.  The authors rely on symbolic computation hence this 
method cannot accept free form surfaces such as splines as input.
In~\cite{sede}, the authors derive a differential equation whose solution is the envelope ${\cal E}$ of the swept volume ${\cal V}$.  
A set of points on ${\cal E}$ is sampled through which a surface is fitted to obtain an approximation of ${\cal E}$.  This method 
accepts smooth solids as input. Further it may not meet the tolerance requirements of modern geometry kernels.  
In~\cite{classifyPoints}, the authors give a complete 
characterization of the points which are inside, outside or on the boundary of the swept volume by giving a point membership test (PMC).
This approach handles class G0 solids, effectively giving a procedural implicit definition (as a PMC) of the swept volume.  Conversion from 
this format to, say, brep format is computationally expensive.
In~\cite{completeSweep}, the authors compute the volume swept by a class G0 cutting tool undergoing 5-axis motion by employing the T-map, i.e., 
the outward normals to the tool at a point.  This method is limited to sweeping tools which are bodies of rotation, and hence, have radial symmetry. 
It does not generalize to sweeping free form solids.
In~\cite{reliable}, the authors present an error-bounded 
approximation of the envelope of the volume swept by a polyhedron along a parametric trajectory.  They employ a volumetric approach using an 
adaptive grid to provide a guarantee about the correctness of the topology of the swept volume.  This approach, however, may not  
meet the tolerance requirements of CAD kernels while being computationally efficient at the same time.
In~\cite{screwMotion}, the authors approximate the given trajectory by a continuous, 
piecewise screw motion and generate candidate faces of the swept surface.  In order to performing trimming, the inverse trajectory method is used.  
Limitations of this method are clear, viz, restriction on the class of motions along which the sweep occurs. 

 In~\cite{sweepComp}, the authors present the first complete computational framework 
for constructing the brep of ${\cal V}$ which is derived from the brep of $M$. Local issues like adjacency 
relations amongst geometric entities of ${\cal E}$ as well as global issues such as their orientation are analysed assuming that $M$ is of 
class G1.  Key constructs such as the {\em prism} and the {\em funnel} are used to parametrize the faces of ${\cal V}$ and guide the 
computation of orientation of co-edges bounding faces of ${\cal V}$.

In this paper we extend the framework proposed in~\cite{sweepComp} to
input solids of class G0.  This, along with the topology and geometry generated 
by smooth faces of $\partial M$ explicated in~\cite{sweepComp} 
and the trimming of swept volume described in~\cite{localGlobal} gives a complete framework for 
computing the brep of the general swept volume.

An edge or a vertex of $\partial M$ is called {\bf sharp} if
it lies in the intersection of faces meeting with G1-discontinuity. 
For instance, in the solid shown in Fig.~\ref{solidFig}, the faces $F_1$ and $F_2$ meet 
in the sharp edge $E_1$ while faces $F_2$ and $F_3$ meet smoothly in edge $E_2$.
The partner co-edges $c_1$ and $c_1'$ for $E_1$ associated with faces $F_1$ and $F_2$ respectively 
and a sharp vertex $Z_1$ are also shown.
While modeling mechanical parts, sharp corners and edges are inevitable features.
Thus this is an important extension of the aforesaid framework.

\begin{figure}
 \centering
 \includegraphics[width=1.0\linewidth]{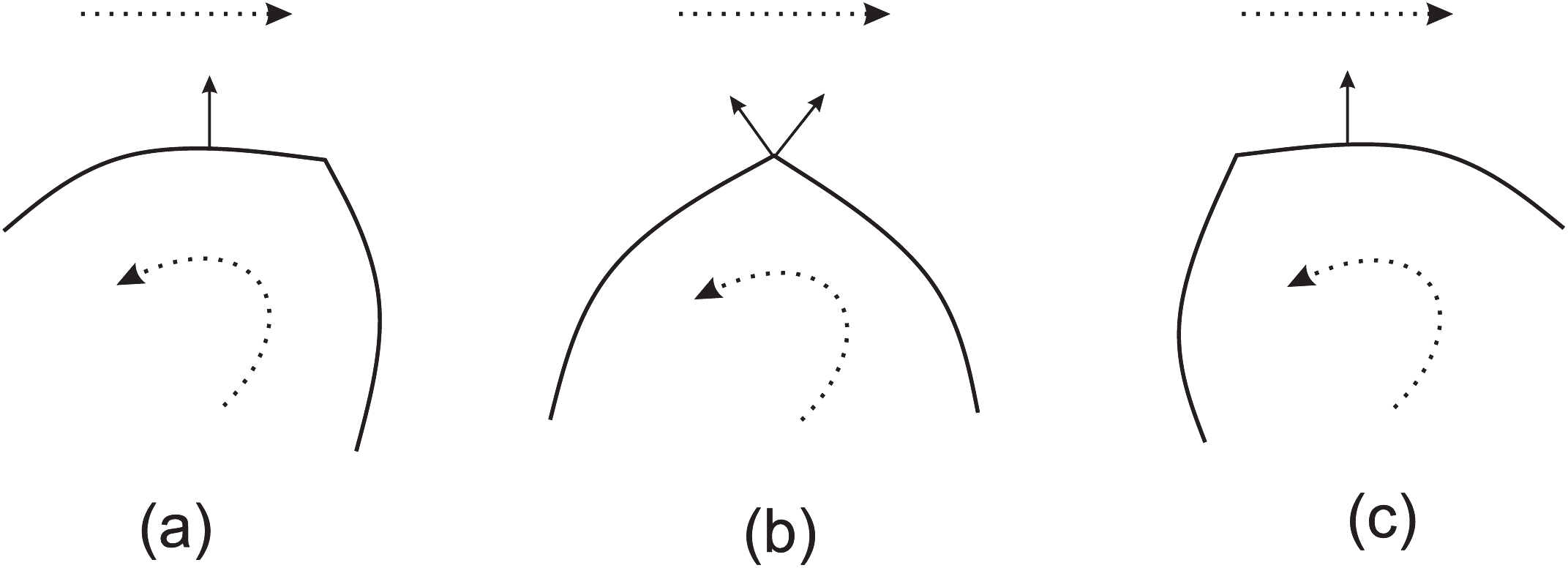}
 \caption{Geometry generated by sharp features.}
 \label{swCurvFig}
\end{figure}

In this work we focus on the entities in the brep of ${\cal E}$ which are generated by sharp edges and 
vertices of $\partial M$.  This involves the following considerations.
\begin{enumerate}
\item 	Geometry: The local geometry of the entity ${\cal E}^E$ in the brep of ${\cal E}$ generated by a sharp edge $E \subset \partial M$ 
can be modeled by that of the 'free' edge $E$ moving in $\mathbb{R}^3$. The surface $S^E$ swept by such an edge is smooth except when 
the velocity at a point is tangent to the edge at that point. 
\item Trim: In order to obtain ${\cal E}^E$, $S^E$ needs to be suitably trimmed.
The correct trimming follows as a result of the interplay between the cone of normals at a sharp point and the trajectory of the point under 
the family of rigid motions.   In the schematic shown in Fig.~\ref{swCurvFig}, an object with sharp features undergoes translation with 
compounded rotation indicated with dotted arrows.  In the positions shown in Fig.~\ref{swCurvFig}(a) and Fig.~\ref{swCurvFig}(c), the sharp 
feature does not generate any points on the envelope while in Fig.~\ref{swCurvFig}(b) it does.  
\item Orientation: The faces ${\cal E}^E$ must be 
oriented so that the unit normal at each point of ${\cal E}^E$ points in the exterior of the swept volume ${\cal V}$.   
\end{enumerate}

We now outline the structure of this paper.  
In Section~\ref{mathStrSec}, we establish a natural correspondence $\pi$ between 
the boundary of the input solid and the boundary of the swept volume. This serves as a basis for a brep structure on ${\cal E}$.  
In Section~\ref{algoSec} we give the overall solid sweep framework and outline 
how it extends the framework proposed in~\cite{sweepComp} to handle sharp features of $\partial M$.
In Section~\ref{calculusSec} we elaborate on the interaction of the unit cone of normals and the trajectory.
In Section~\ref{paramSec} we parametrize 
the faces and edges of ${\cal E}$ generated by sharp features of $\partial M$. 
In Section~\ref{topoSec} we analyse the  adjacency relations amongst 
of entities of ${\cal E}$ via the correspondence map $\pi$.  We show that there is local similarity between 
the brep structure of ${\cal E}$ and that of $\partial M$. 
In Section~\ref{brepSec} we explain the steps of the overall computational framework given in Section~\ref{algoSec}.
We give many sweep examples demonstrating the effectiveness of our algorithm.
In Section~\ref{extensions}, we discuss subtle issues of self-intersections and how they can be handled.
Finally, we conclude in Section~\ref{concludeSec} with remarks on extension of this work.

\section{Mathematical structure of the sweep}	\label{mathStrSec}

In this section we define the envelope obtained by sweeping the input solid $M$ along 
the given trajectory $h$.  

\begin{defn} \label{trajectoryDef}
A {\bf trajectory} in $\mathbb{R}^3$ is specified by a map 
\begin{align*}
h:I \rightarrow (SO(3), \mathbb{R}^3), h(t) = (A(t), b(t))
\end{align*}
where $I$ is a closed interval of $\mathbb{R}$, $ A(t) \in SO(3) \footnote{$SO(3)=\{X \mbox{ is a 3 $\times$3 real matrix} |X^t \cdot X = I, det(X)=1  \}$ 
is the special orthogonal group, i.e. the group of rotational transforms.}, b(t) \in \mathbb{R}^3$.    The parameter $t$ represents time.    
\end{defn}

We assume that $h$ is of class $C^k$ for some $k \geq 2$, i.e., partial derivatives of order up to $k$ exist and are continuous. 

We make the following key assumption about $(M,h)$.

\begin{assum} \label{genericAssum}
The tuple $(M,h)$ is in a {\em general position}.
\end{assum}

\begin{defn}  \label{envlDef}
The {\bf action} of $h$ (at time $t$ in $I$) on $M$ is given 
by $M(t) = \{ A(t) \cdot x + b(t) | x \in M\}$.  
The {\bf swept volume} ${\cal V}$ is the union 
$ \bigcup_{t \in I} M(t)$ and the {\bf envelope} ${\cal E}$ is defined as the 
boundary of the swept volume ${\cal V}$.  
\end{defn}

\begin{figure}
 \centering
 \includegraphics[width=1.0\linewidth]{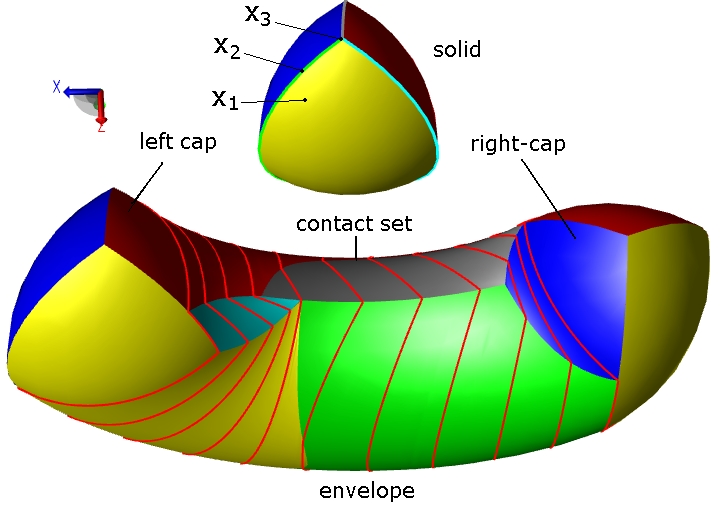}
 \caption{A solid undergoing translation along a circular arc in $xy$-plane and rotation about $y$-axis. 
Curves of contact at few time instants are shown on the envelope in red.}
 \label{contactSetFig}
\end{figure}

An example of a swept volume appears in Fig.~\ref{contactSetFig}.
Clearly, for each point $y$ of ${\cal E}$ there must be an $x \in M$ and a $t \in I$ 
such that $y = A(t) \cdot x + b(t)$. 

We denote the interior of a set $W$ by $W^o$ and its boundary by $\partial W$.
It is clear that ${\cal V}^o = \cup_{t \in I} M(t)^o$. Therefore, 
if $x \in M^o$, then for all $t \in I$,  $A(t) \cdot x + b(t) \notin {\cal E}$.
Thus, the points in the interior of $M$ do not contribute any point on the envelope.

\begin{defn} \label{trajXDef}
For a point $x \in M$, define the {\bf trajectory of} $\pmb{x}$ as the map $\gamma_x : I \to \mathbb{R}^3$ 
given by $\gamma_x(t) = A(t) \cdot x + b(t)$ and the velocity $v_x(t)$ as 
$v_x(t) = \gamma_x'(t) = A'(t) \cdot x + b'(t)$.
\end{defn}

Now we recall the fundamental proposition (\cite{sede, sweepComp}) which assumes that $M$ is {\em smooth} and
provides a necessary condition for a point $x \in \partial M$ to contribute the point $\gamma_x(t)$
on ${\cal E}$ at time $t$.

\begin{prop}\label{smoothprop} Let $M$ be smooth and for $x \in \partial M$, let $N_x$ be the unit
outward normal to $M$ at $x$.
Define the function $G: {\partial M} \times I \rightarrow \mathbb{R}$
as $ G(x,t) = \left< A(t) \cdot N_x, v_x(t) \right> $. In other words, $G(x, t)$ is the dot product of the velocity 
vector with the unit outward normal at the point $\gamma_x(t) \in \partial M(t)$.

Further, let $I = [t_0, t_1], t \in I$ and $x \in \partial M$ be such that $\gamma_x(t) \in \mathcal{E}$. Then either 
(i) $G(x,t) = 0$ or 
(ii) $t = t_0$ and $G(x,t) \leq 0$, or 
(iii) $t = t_1$ and $G(x,t) \geq 0$.
\end{prop}

Now we develop some notation in order to generalize the above proposition to non-smooth $M$ represented in the brep
format. Recall that the brep of $M$ models $\partial M$ through a collection of faces which meet each other
across edges which in turn meet at vertices. Clearly, the sharp features of $M$ are located along the edges and
vertices.

\begin{figure}
 \centering
 \includegraphics[scale=0.3]{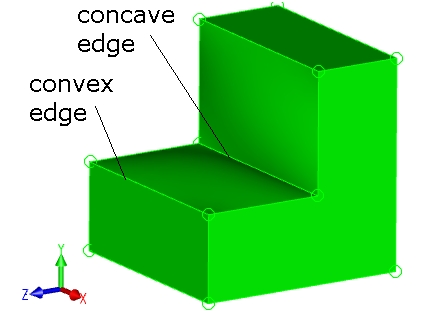}
 \caption{Convex and concave edges on a solid.}
 \label{concaveFig}
\end{figure}

The solid $M$ may be (partly) convex/concave at a sharp edge.  See Fig.~\ref{concaveFig} for an example.
\eat{Fig.~\ref{concaveFig} shows a solid with three concave edges.}  
For the moment we only consider solids that do not have concave edges.  See Section~\ref{extensions} for a discussion on concave edges.
\eat{It is easily shown that concave edges do not give rise to any geometric entity on 
the envelope.}
Further, for simplicity, we assume that at most three faces meet at a sharp vertex in $\partial M$.
\eat{Consider a sharp point $x \in \partial M$ ($x$ is either a vertex or an interior point of an edge) which lies in the intersection of faces $F_i$ for 
$i = 1, \ldots, m$, $m\leq 3$. There exists a cone of {\em outward} normals at $x$, namely the cone
formed by $N_i, i = 1, \ldots, m$, where $N_i$ is the unit outward normal to $F_i$ at $x$.}

\eat{
\begin{defn} 	\label{coneDef}
For a point $x \in  \bigcap_{i=1}^{m} F_i$, define the {\bf cone of unit normals} at $x$ as 
${N}_x = \left  \{ \displaystyle \frac{  \sum_{i=1}^m \alpha_i \cdot N_i }{ \left \|  \sum_{i=1}^m \alpha_i \cdot N_i \right \| } \right \}$, 
where, $N_i$ is the unit outward normal to face $F_i$  
at point  $x$  and  $\alpha_i \in \mathbb{R}, \alpha_i \geq 0$ for $i = 1, \ldots, m$  and  
$ \sum_{i=1}^{m} \alpha_i= 1 $.  
\end{defn}
}

\begin{defn}	\label{coneDef}
For a point $x \in \bigcap_{i=1}^{m} F_i$, define the {\bf cone of unit (outward) normals} (to $\partial M$) at $x$ as 
the intersection of the unit sphere $S^2$ with the cone formed by $N_i$, for $i = 1, \ldots, m$, where
$N_i$ is the unit outward normal to $F_i$ at $x$.
For simplicity, we assume that $N_i$ for $i= 1, \ldots, m$ are linearly independent. 
We denote the cone of unit normals at $x$ by $N_x$. 
\end{defn}

The points labeled $x_3$ and $x_2$ in Fig.~\ref{contactSetFig} 
lie in the intersection of three and two smooth faces respectively meeting sharply.
The point labeled $x_1$ lies in the interior of a smooth face, hence $N_{x_1}$ has 
a single element, namely, outward normal to $\partial M$ at $x_1$.
The cone of normals is referred to as the {\em extended Tool map} in~\cite{completeSweep}.

\eat{Note that the cone of unit normals in Definition~\ref{coneDef} applies only to convex edges.}

\begin{figure}
 \centering
 \includegraphics[scale=0.4]{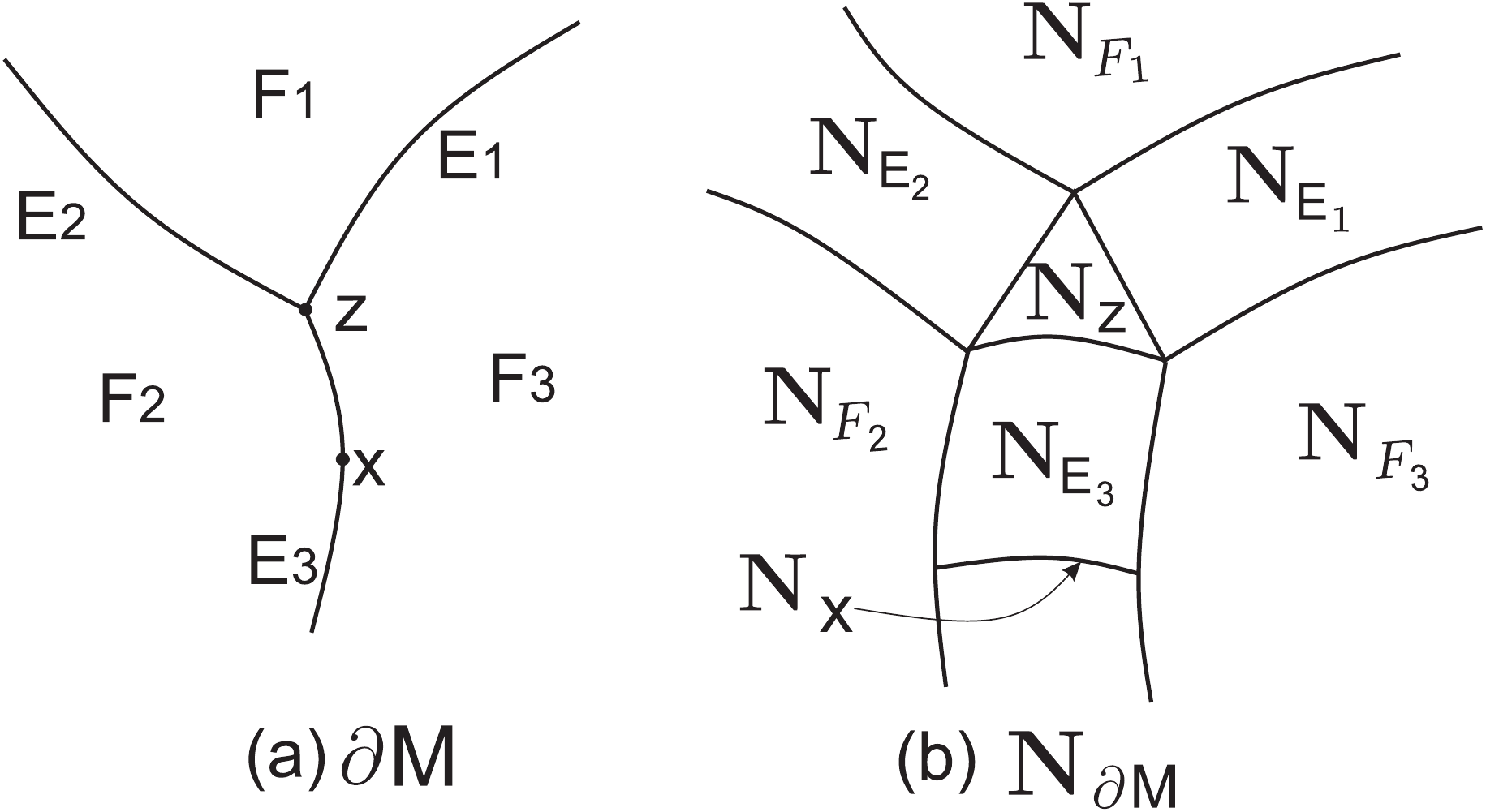}
 \caption{A solid and its unit normal bundle.}
 \label{coneFig}
\end{figure}

\begin{defn}	\label{coneBunDef}
For a subset $X$ of $\partial M$, the {\bf unit normal bundle} (associated to $X$) is defined as the disjoint union of the cones of unit normals at each point 
in $X$ and denoted by ${\bf N_X}$, i.e., ${\bf N_X} =  \bigsqcup_{x \in X} {N}_x =  \bigcup_{x \in X}\{ (x, n) | n \in { N}_x \}$.  
\end{defn}

 In Fig.~\ref{coneFig}(a) a portion of $\partial M$ is shown in which three faces $F_i$ and three edges 
$E_i$ meet at a sharp vertex $Z$.  Note that for $X \subset \partial M$, 
${\bf N_X} \subset \mathbb{R}^3 \times S^2$, where $S^2$ is the unit sphere in $\mathbb{R}^3$.
However, for the ease of illustration we have shown the unit normal bundles ${\bf N_{F_i}}, {\bf N_{E_i}}$ for $i = 1, 2, 3$ and 
${\bf N_Z}$ schematically in Fig.~\ref{coneFig}(b) in which an element $(x, n) \in {\bf N_{\partial M}}$ is 
represented as the `offset' point $x + n$.


For $x \in \partial M$ and $t \in I$, the cone of unit normals to $\partial M(t)$ at the point $\gamma_x(t)$ is 
given by $A(t) \cdot {N}_x := \{ A(t) \cdot n  | n \in {N}_x \}$.  Further, the action of $h$ 
at time $t \in I$ on the unit normal bundle ${\bf N_{\partial M}}$ is given by 
${\bf N_{\partial M}}(t) := \{ (\gamma_x(t), A(t) \cdot n) | x \in \partial M, n \in {N}_x \}$.

\begin{defn}	\label{gDef}
For  $(x, n) \in {\bf N_{\partial M}}$ and $t \in I$, define the function 
$g:{\bf N_{\partial M}} \times  I \to \mathbb{R}$ as 
$ g(x,n,t) = \left < A(t) \cdot n , v_x(t) \right > $.
\end{defn} 
Thus, $g(x,n, t)$ is the dot product of the velocity with the normal $A(t) \cdot n \in A(t) \cdot { N}_x$ 
at the point $\gamma_x(t) \in \partial M(t)$.

We are now ready to state the next Proposition which is a natural generalization of Proposition~\ref{smoothprop} to non-smooth solids.

\begin{prop}	\label{gProp}
Let $I = [t_0, t_1], t \in I$ and $x \in \partial M$ be such that $\gamma_x(t) \in \mathcal{E}$. Then either 
(i) $t = t_0$ and there exists $n \in {N}_x$ such that $g(x, n, t) \leq 0$ or 
(ii) $t = t_1$ and there exists $n \in {N}_x$ such that $g(x, n, t) \geq 0$ or
(iii) There exists $n \in {N}_x$ such that $g(x, n, t) = 0$.
\end{prop}
For proof refer to Appendix~\ref{proof1Sec}.

\begin{defn}	\label{cocDef}
Fix a time instant $t \in I$. The set 
$\{ \gamma_x(t) \in \partial M(t)| \exists n \in {N}_x \mbox{ such that } g(x, n,t) = 0 \}$ 
is referred to as the {\bf curve of contact} at $t$ and denoted by $C(t)$.  
The set $\{ (\gamma_x(t),  A(t) \cdot n) \in {\bf N_{\partial M}}(t) | g(x, n, t) = 0 \}$ 
is referred to as the {\bf normals of contact at $t$} and denoted by ${\bf C}(t)$.  Further, 
the union of curves of contact is referred to as the {\bf contact set} and denoted by $C$,
i.e., $C =  \bigcup_{t \in I} C(t)$.
The union $\bigcup_{t\in I} {\bf C}(t)$ is 
referred to as the {\bf normals of contact} and denoted by ${\bf C}$. 
\end{defn}

\begin{figure}
 \centering
 \includegraphics[scale=0.45]{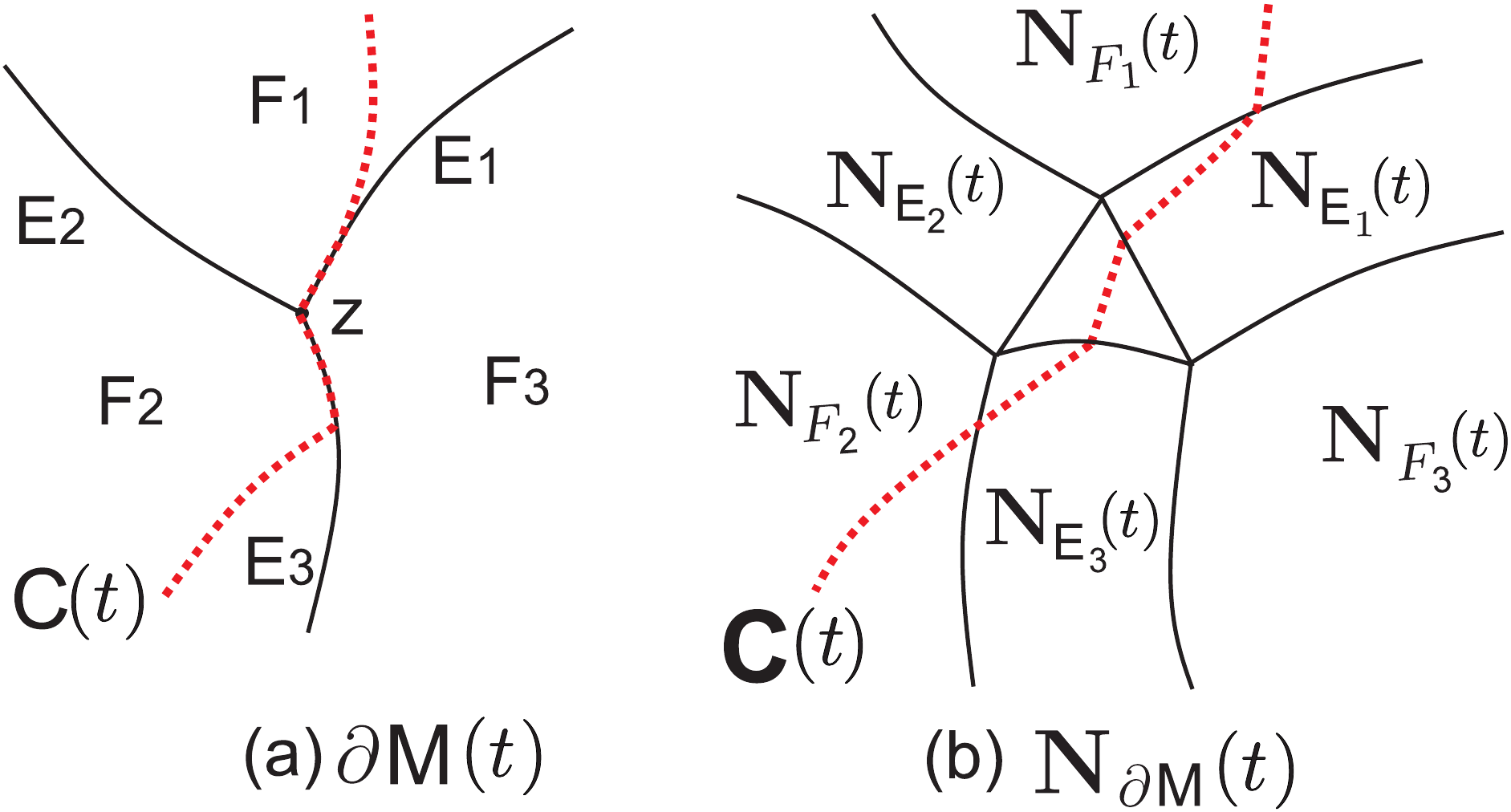}
 \caption{Curve of contact and normals of contact at time $t$}
 \label{cocFig}
\end{figure}

Curves of contact at a few time instants are shown in the sweep example of Fig.~\ref{contactSetFig} in red.
Fig.~\ref{cocFig} schematically illustrates the normals of contact and the curve of contact at a time 
instant $t$ shown as dotted curves in red. 
The curve of contact is referred to as the characteristic curve in~\cite{peternell}.
The normals of contact at $t$ are referred to as the contact map in~\cite{completeSweep}.

The {\bf left cap} is defined as $L_{cap}=\{ \gamma_x(t_0) \in \partial M(t_0)| \exists n \in N_x \mbox{ such that } g(x,n,t_0) \leq 0 \}$ and 
the {\bf right cap} is defined as $R_{cap}=\{ \gamma_x(t_1) \in \partial M(t_1)| \exists n \in N_x \mbox{ such that } g(x,n,t_1) \geq 0 \}$.  
The left cap and right cap are shown in the sweep example of Fig.~\ref{contactSetFig}.
The left and right caps can be easily computed from the solid at initial and final positions.  

Note that, by Proposition~\ref{gProp}, ${\cal E} \subseteq L_{cap} \cup C \cup R_{cap}$. 
In general, a point on the contact set $C$ may not appear on the complete envelope ${\cal E}$ as it may get occluded
by an interior point of the solid at a different time instant, see for example Fig.~\ref{toolFig}.
 This complicates the correct construction
of the envelope by an appropriate {\em trimming} of the contact-set. We refer the reader to \cite{localGlobal} for a 
comprehensive mathematical analysis of the trimming and the related subtle issues arising due to local/global intersections 
of the family $\{C(t)\}_{t \in I}$. In this paper, we focus on the case of {\em simple} sweeps.

\begin{defn} 	\label{simpleDef}
A sweep $(M, h)$ is said to be simple if the envelope is the union of the contact set, the left cap and the right cap, i.e.,  
${\cal E} = L_{cap} \cup C \cup R_{cap}$.
\end{defn}
Hence, in a simple sweep, every point on the contact set appears on the envelope and no trimming of the contact set is 
required in order to obtain the envelope.

\begin{lem}		\label{cocIntLem}
For a simple sweep, for $t \neq t'$, $C(t) \cap C(t') = \emptyset$.  In short, no two distinct curves of contact intersect.
\end{lem}
Refer to~\cite{sweepComp} for proof.
\begin{defn} 	\label{piDef}
For a simple sweep, define the natural correspondence 
$\pi: C \to \partial M$ as follows: for $y \in C(t)$, we set $\pi(y)=x$ where $x$ is the unique point on $\partial M$
such that $\gamma_x(t)=y$. 
\end{defn}
Thanks to Lemma~\ref{cocIntLem}, $\pi$ is well-defined. Thus, $\pi(y)$ is the natural point on $\partial M$
which transforms to $y$ through the sweeping process.

Further, define the natural `normal' correspondence
${\pmb \pi} : {\bf C} \to {\bf N_ {\partial M}}$ 
as ${\pmb \pi}((y,n)) = (x, n')$ if $(y,n) = (\gamma_x(t), A(t) \cdot n')$ for the unique $t \in I$ and the unique 
$(x, n') \in {\bf N_{\partial M}}$ such that $g(x,n',t)=0$ (cf. Proposition~\ref{gProp} and Definition~\ref{cocDef}).

\begin{figure}
 \centering
\includegraphics[scale=0.35]{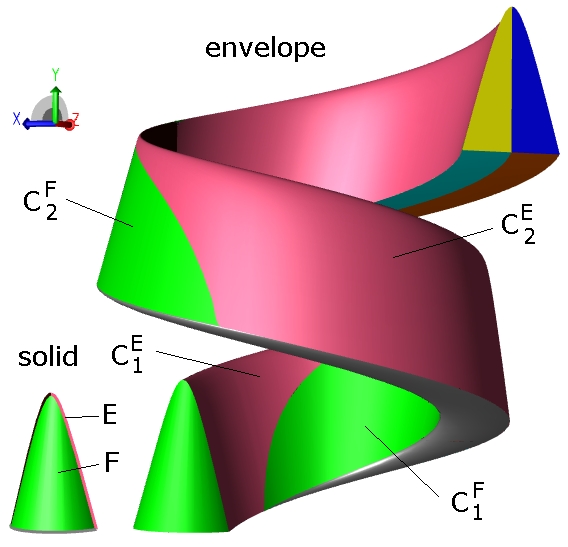}
 \caption{The sharp edge $E \subset \partial M$ generates two faces $C^E_1$ and $C^E_2$, all shown in pink.  
The face $F \subset \partial M$ generates two faces $C^F_1$ and $C^F_2$, all shown in green.}
 \label{coneCubeHelixFig}
\end{figure}

The correspondence $\pi$ induces a natural brep structure on ${\cal E}$ which is derived from that of $\partial M$.
The map $\pi$ is illustrated via color coding in the sweep examples shown in Figures~\ref{contactSetFig},~\ref{coneCubeHelixFig},~\ref{singExFig} 
and~\ref{showCaseFig} by showing the points $y$ and $\pi(y)$ in the same color.

\eat{
\begin{figure}
 \centering
 \includegraphics[scale=0.4]{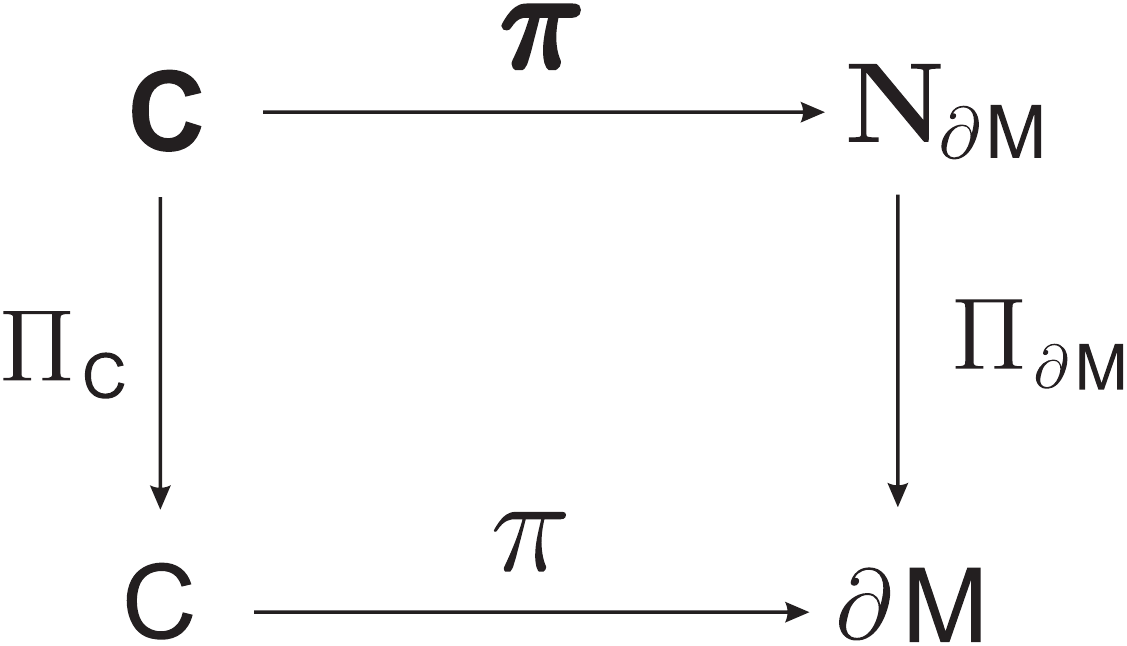}
 \caption{The above diagram commutes.}
 \label{piFig}
\end{figure}
}

A face of $\partial M$ generates a set of faces on the contact set $C$. An edge or a vertex 
where $\partial M$ is G1-continuous generates a set of edges or vertices respectively on $C$.
In other words, a G1-continuous subset $O$ of $\partial M$ generates entities on $C$ whose 
dimension is same as that of $O$.  In the sweep example shown in Fig.~\ref{coneCubeHelixFig}, the 
face $F \subset \partial M$ generates a set of faces on $C$.
 However, a sharp edge of $\partial M$ generates a set of 
faces on $C$ and a sharp vertex generates a set of edges on $C$. 
This is illustrated in the example of Fig.~\ref{coneCubeHelixFig} by the sharp 
edge labeled $E \subset \partial M$ which generates faces on $C$ shown in pink.
For $O \subseteq \partial M$, we denote the contact set generated by $O$ by $C^O$, i.e., 
$C^O = \{ y \in C | \pi(y) \in O\}$. Note that while $O$ is connected, the 
corresponding contact set $C^O$ may not be. 
A connected component of $C^O$ is denoted using 
a subscript, for example, faces $C^E_1$ and $C^E_2$ in Fig.~\ref{coneCubeHelixFig} 
correspond to the edge $E \subset \partial M$.


\section{The computational framework} 	\label{algoSec}

Algorithm~\ref{frameworkAlgo} given below provides an outline of the basic Algorithm 1
of~\cite{sweepComp} and its extension to sharp edges, which begins on Step 14, and
which is the main contribution of the paper.

\begin{algorithm}
\caption{Solid sweep} \label{frameworkAlgo}
\begin{algorithmic}[1]
\ForAll {faces $F$ in $\partial M$}
	\ForAll {co-edges $c$ in  $\partial F$}
		\ForAll {$z$ in $\partial c$}
 			\State Compute vertices $C^z$  generated by $z$
		\EndFor
		\State Compute co-edges $C^c$ generated by $c$
		\State Orient co-edges $C^c$
	\EndFor
	\State Compute $C^F(t_0)$ and $C^F(t_1)$
	\State Compute loops bounding faces $C^F$ which will be generated by $F$
	\State Compute faces $C^F$ generated by $F$
	\State Orient faces $C^F$
\EndFor
\ForAll {sharp edges $E$ in $\partial M$}
	\ForAll {$Z$ in $\partial E$}
		\State Compute co-edges $C^Z$ generated by $Z$
		\State Orient co-edges $C^Z$
	\EndFor
	\State $(F, F') \leftarrow AdjacentFaces(E)$
	\State Compute co-edges $C^E \cap C^F$ and $C^E \cap C^{F'}$
	\State Orient co-edges $C^E \cap C^F$ and $C^E \cap C^{F'}$
	\State Compute $C^E(t_0)$ and $C^E(t_1)$
	\State Compute loops bounding faces $C^E$ which will be generated by $E$
	\State Compute faces $C^E$ generated by $E$
	\State Orient faces $C^E$
\EndFor
\State Compute adjacencies between faces of $C$
\end{algorithmic}
\end{algorithm}

We outline what was achieved in~\cite{sweepComp}. At the heart of Algorithm 1 is an
entity-wise {\em implementation} of the correspondence $\pi $ which is a 
classification of the faces, edges and vertices of ${\cal E}$ by the
generating entity in $\partial M$. This is achieved by computing $C^O $ of the envelope
for key entities $O \subseteq \partial M$ which yield faces in ${\cal E}$. The smooth case is easy since faces
generate faces, edges generate edges and so on. The computation of $C^O$ is
followed by an orientation calculation. It was noted that while the
adjacencies of entities in ${\cal E}$ were built from that on $\partial M$, the
orientation on ${\cal E}$ was {\em not} as on $\partial M$ and in fact
could be positive, negative or zero {\em vis a vis} that on $\partial M$.

Let us outline the details of the computation of $C^F$ for a smooth face $F \subseteq \partial M$. 
Suppose that $F$ is given by the parametrization $S:D \rightarrow \mathbb{R}^3$, where $D$ is a domain in
$\mathbb{R}^2 $ with parameters $(u,v)$. Let $I$ be the interval used to parametrize
the motion $h$. The envelope condition (cf. Proposition~\ref{smoothprop}) yields 
a function $f^F(u,v,t)$ on
the {\em prism} $D \times I$, viz., $f^F(u,v,t)=\langle A(t)\cdot N(u,v) ,\gamma_{S(u,v)}'(t)
\rangle$ where $N(u,v)$ is the outward normal to $F$ at $S(u,v)$. 
For simple sweeps $f^F(u,v,t)=0$ indicates that $A(t)\cdot S(u,v) +
b(t)$ is on the envelope. This led to the definition of the funnel
${\cal F}^F$ as the zero-set of $f^F$ within the prism. If
${\cal F}^F_1 , \ldots, {\cal F}^F_k $ are the connected components of the funnel
then (i) the face $F$ leads to exactly $k$ disjoint faces $C^F_1 ,\ldots ,
C^F_k$ in the envelope ${\cal E}$, (ii) each ${\cal F}^F_i $ serves as the
parameter space to implement $C^F_i $, (iii) the boundary of
${\cal F}^F_i $ arises from ${\cal F}^F$ intersecting the boundary of the prism
and parametrizes the co-edges of $C^F_i $. The above computation is achieved
in Steps 1 to 13 of Algorithm~\ref{frameworkAlgo}.

\begin{figure}
 \centering
 \includegraphics[scale=0.4]{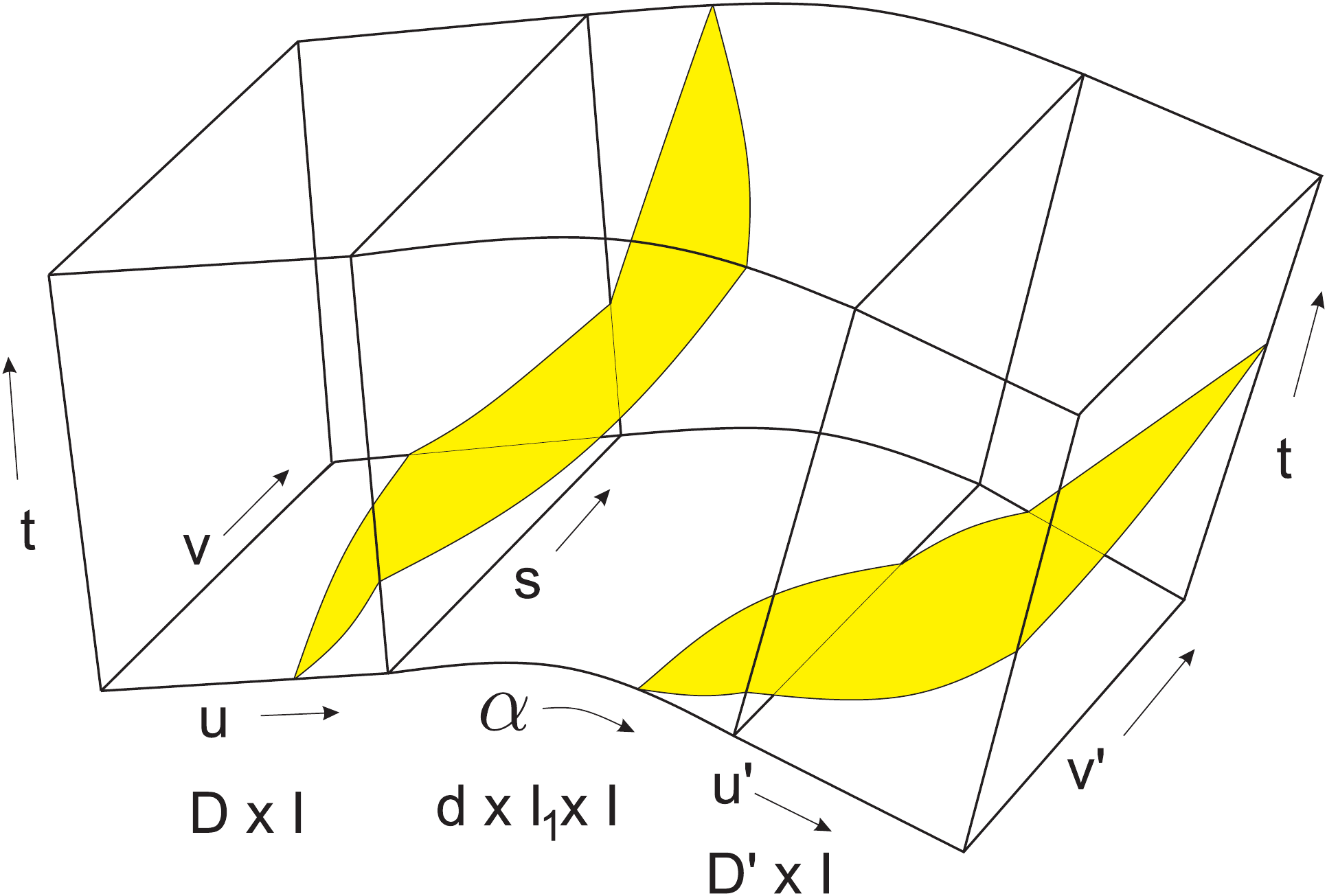}
 \caption{Prisms for faces $F, F'$ and edge $E$ shown adjacent to each other.  The funnels of $F, F'$ and pre-funnel of $E$ are shaded in yellow.}
 \label{prismFig}
\end{figure}

The same approach works when the solid has sharp features, albeit with some
complications. Firstly, a sharp edge generates a {\em face} and a sharp
vertex an {\em edge}. This is because, for a point $x$ on a sharp
edge, there is actually a {\em cone of normals} $N_x $ (cf. Definition~\ref{coneDef}). Whence $\gamma_x(t)$
is on the envelop ${\cal E}$ iff the velocity $\gamma_x' (t)$ is perpendicular
to any element of $A(t).N_x$ (cf. Proposition~\ref{gProp}). Thus, this results in the sharp edge $E$ in extruding
a 2-dimensional entity. The analysis of the smooth face via the prism and
the funnel lifts easily and naturally to the case of the sharp edge $E$. The
prism is ${\bf N_E} \times I$, suitably parametrized, which is a $3$-dimensional
entity. The envelope condition leads to an implicit surface {\em pre-funnel}.
The funnel ${\cal F}^E $ is the projection of the pre-funnel on to
$E \times I$.
Thus, for a point $x \in E$ and $t\in I$, if $(x,t)\in {\cal F}^E $, then
$\gamma_x(t)$ is on the envelope. See Fig.~\ref{prismFig} for an illustration of how
funnels of smooth faces interact with the pre-funnel of the sharp edge.

The geometry of $C^E $ is simpler: it is merely the geometry of a translate/sweep
of a curve and is implemented routinely in most kernels. Further, the orientation
of a face of $C^E$ is also shown to be easily computable.

The trims/boundary of a face of $C^E $ is obtained by examining the components of ${\cal F}^E$ whose
boundaries are shown to be intimately related to the normal cones.
Next, for a sharp vertex $Z$, it is easy to compute a set of sub-intervals of
$I$ when appropriate translates of $Z$ will appear {\em as edges} on ${\cal E}$.
The computation of adjacencies between the new entities is governed by a simple yet rich interplay between the 
normal cones at sharp points and their trajectories. 

The key technical contributions thus are essentially (i) a calculus of the sweep of normal
cones and its embedding into a brep framework (ii) a seamless architectural integration of 
sharp features into the general solid sweep framework.
An obvious question is why
it could not have been done before, i.e., in~\cite{sweepComp} itself. The answer is of
course that the structure of sweeps $C^F $ of smooth faces is the key
construct and the $C^E $, i.e., sweeps of sharp edges are essentially {\em transition
faces}. Thus the theory of these transition faces must be subsequent
to that of the smooth faces.

\section{Calculus of cones}	\label{calculusSec}

In this section we develop the mathematics of the interaction between the
cones at sharp points and their trajectories under $h$. 

Towards this, fix a sharp point $x$ with normal cone $N_x$ and a time instant $t$.
Proposition~\ref{gProp} provides a geometric condition which determines if
$\gamma_x(t)$ will be on ${\cal E}$. Namely, $\gamma_x(t) \in {\cal E}$ iff there
exists $n \in N_x$ such that $g(x,n,t)=\left< \gamma_x'(t), A(t).n \right> = 0$.

Further, if $n_1, n_2 \in N_x$ are such that $g(x,n_1,t)=g(x,n_2,t)=0$ then for any {\em linear} combination $n \in N_x$ of $n_1$ and $n_2$,
$g(x,n,t)=0$.
This follows by observing that, having fixed $x$ and $t$, the function $g(x,n,t)$ is linear in $n$.

\eat{
Furthermore, the following lemma characterizes the nature of the normals of contact as in Definition~\ref{cocDef}.
\begin{lem}\label{linearity} Let $x \in \partial M$ with normal cone $N_x$ and a time instant $t$ so that $\gamma_x(t) \in {\cal E}$.
If $n_1, n_2 \in N_x$ are such that $g(x,n_1,t)=g(x,n_2,t)=0$. Then for any {\em linear} combination $n \in N_x$ of $n_1$ and $n_2$,
$g(x,n,t)=0$.
\end{lem}
\noindent {\em Proof.}
The lemma follows by simply observing that, having fixed $x$ and $t$, the function $g(x,n,t)$ is linear in $n$.
\hfill $\square$
}

\eat{
For a sharp point $x$ of $\partial M$, if $\left < A(t) \cdot n, \gamma_x'(t) \right > = 0$ and $\left < A(t) \cdot n', \gamma_x'(t) \right >= 0$ 
for $n, n' \in N_x$, $n \neq n'$ then $\left <\alpha_1 \cdot A(t) \cdot n + \alpha_2 \cdot A(t) \cdot n', \gamma_x'(t) \right > = 0$ 
for all $\alpha_1, \alpha_2 \in \mathbb{R}$.
\end{lem}

In other words for $\gamma_x(t) \in C(t)$, the set $\{ n \in N_x | \left < A(t) \cdot n , \gamma_x'(t) \right > = 0 \}$ is either 
singleton or of infinite cardinality.

This is done in two
steps, viz., (i) for a point $x$ either on
an edge or a vertex, with normal cone $N_x $, and a time instant $t$, we
derive clean geometric conditions which determine if $\gamma_x (t)$ will
be on ${\cal E}$, and next (ii) we prove that when $\gamma_x (t)$ is on
${\cal E}$, it is benign.
}

\subsection{Interaction between $N_x$ and $\gamma_x(t)$ on a sharp edge}	\label{coneSec}

Consider a sharp edge $E$ bounding the smooth faces $F, F'$ in $\partial M$.
Further, fix an interior point $x$ on $E$ and a time instant $t$.
Let $N_1$ and $N_2$ be the unique unit outward normals to $F$ and $F'$ at $x$.  
Note that that the normal cone $N_x$ is `spanned' by $N_1$ and $N_2$.

\begin{figure}
 \centering
\includegraphics[scale=0.5]{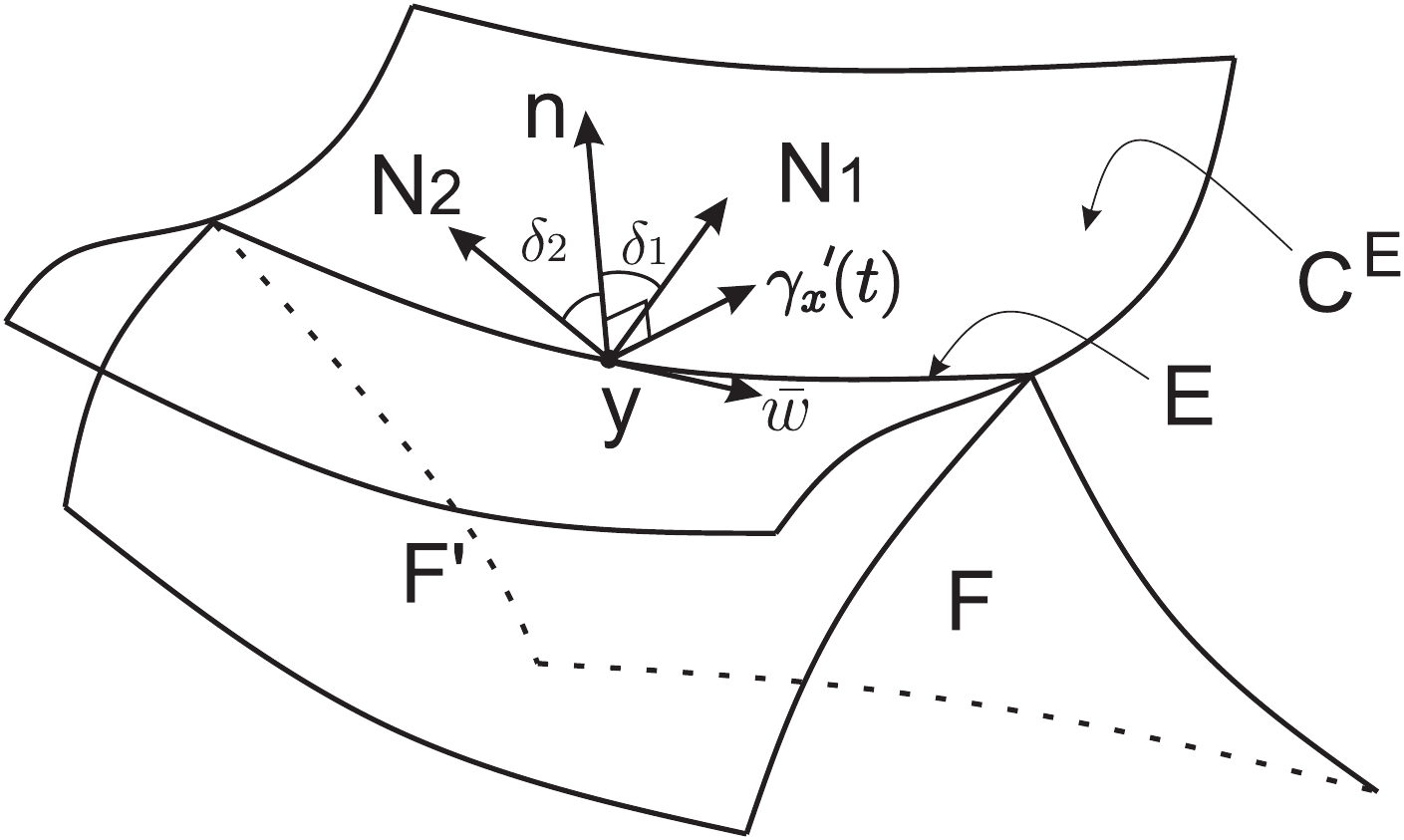}
 \caption{The cone of unit normals at $N_x$ formed by normals $N_1$ and $N_2$ to faces $F$ and $F'$ respectively meeting in sharp edge $E$.}
 \label{normalFig}
\end{figure}

Let $\bar{w}$ be the tangent to $E$ at $x \in E$. Clearly, for every $n \in N_x$, 
$\left < n, \bar{w} \right > = 0$ and thus, $\left< A(t)\cdot n, A(t) \cdot \bar{w} \right>=0$.
Now $\left < \gamma_x'(t), A(t) \cdot n \right > = 0$ 
for some $n \in N_x$ iff $A(t) \cdot n$ is parallel to $A(t) \cdot \bar{w} \times \gamma_x'(t)$. 
Hence, $\gamma_x(t) \in {\cal E}$ if and 
only if $A(t) \cdot \bar{w} \times \gamma_x'(t) \in A(t) \cdot N_x$ or 
$-A(t) \cdot \bar{w} \times \gamma_x'(t) \in A(t) \cdot N_x$. This is illustrated 
schematically in Figure~\ref{normalFig}. 

\eat{We later show (cf. Section~\ref{orientCESec}) that this condition is intimately
related to the orientation of $C^E$.} 
Further, note that, if $\gamma_x(t) \in {\cal E}$ then we have the following
dichotomy: either there exists a {\em unique} $n \in N_x$ such that $g(x,n,t)=0$ or for {\em all} $n \in N_x$ $g(x,n,t)=0$.
It is easy to see that the later condition is equivalent to
$A(t) \cdot \bar{w} \times \gamma_x'(t)=\bar{0}$ and as shown in Section~\ref{singSec} leads
to a singularity on $C^E$.

\begin{figure}
 \centering
\includegraphics[scale=0.7]{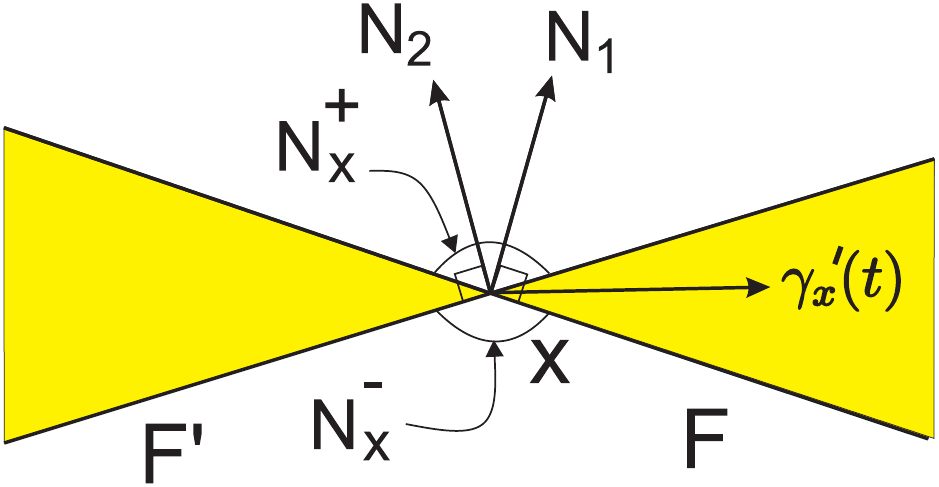}
 \caption{The point $\gamma_x(t)$ is on contact set if and only if $\gamma_x'(t)$ is in the region shaded in yellow.}
 \label{candyFig}
\end{figure}

For further discussion, we assume without loss of generality that $A(t) = I$ and $b(t) = 0$.
Define $N_x^- = \{ \bar{z} \in \mathbb{R}^3 | \left < n , \bar{z} \right > < 0 \  \forall n \in N_x \}$ 
and $N_x^+ = \{ \bar{z} \in \mathbb{R}^3 | \left < n , \bar{z} \right > > 0 \  \forall n \in N_x \}$.  
$N_x^-$ and $N_x^+$ are illustrated schematically in Figure~\ref{candyFig}.  By Proposition~\ref{gProp}, 
$\gamma_x'(t) \in N_x^+ \cup N_x^-$ iff $\gamma_x(t) \notin {\cal E}$. The complement of $N_x^+ \cup N_x^-$ 
is shaded in yellow in Figure~\ref{candyFig}.  It is easy to see that, $\gamma_x(t) \in {\cal E}$ if and only if either 
(i) $\left < N_1, \gamma_x'(t) \right > \leq 0$ and $\left < N_2, \gamma_x'(t) \right > \geq 0$
or (ii) $\left < N_1, \gamma_x'(t) \right > \geq 0$ and $\left < N_2, \gamma_x'(t) \right > \leq 0$.
This condition is computationally easy to check and is used to define the trim curves of $C^E$.

\eat{Besides this, membership in $N_x^+$ (respectively, $N_x^-$) also defines the orientation of 
$C^E$ (cf. Section~\ref{orientCESec}).}

\subsection{Interaction between $N_x$ and $\gamma_x(t)$ at a sharp vertex}

Consider now the case when $x$ is a vertex with face normals $N_1, N_2 ,N_3 $ coming from faces $F_1 ,F_2 ,F_3 $ respectively. 
As before, for simplicity, we assume that $A(t)=I$ and $b(t)=0$.

\begin{figure}
 \centering
\includegraphics[scale=0.42]{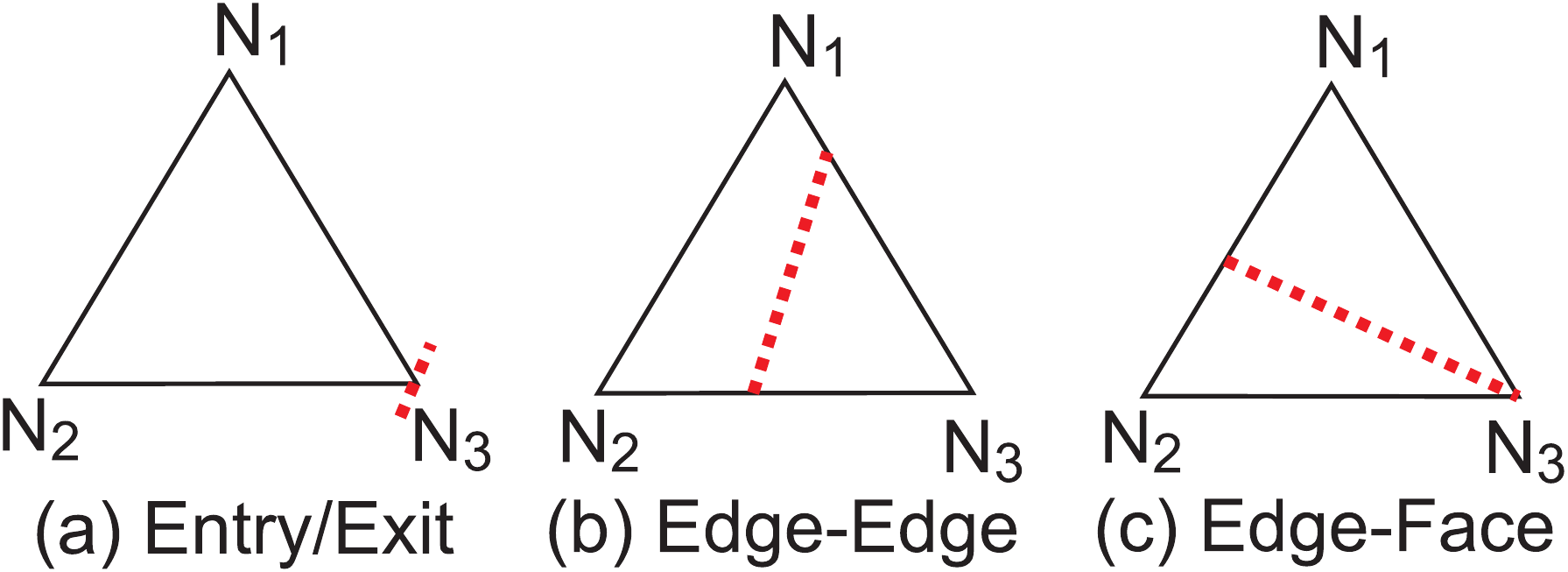}
 \caption{Three possible configurations of the normals of contact at $t$ for sharp vertex $Z$.  
The set $\{ n \in N_Z| \left < n , \gamma_z'(t) \right > = 0\}$ is shown as a dotted curve in red in ${\bf N_Z}$.}
 \label{adjConfigFig}
\end{figure}

Once again, $\gamma_x(t) \in {\cal E}$ iff there is an $n \in N_x $ such that $\langle n , \gamma_x'(t)\rangle = 0$.
Figure~\ref{adjConfigFig} schematically illustrates the set
$\{ n \in N_x| \left < n , \gamma_x'(t) \right > = 0\}$ of normals of contact at time $t$. 
An important observation is that this set is closed under linear combinations. Therefore, upto
permutations of $N_i$'s, Figure~\ref{adjConfigFig} describes all the configurations which
lead to $\gamma_x(t) \in {\cal E}$.

It is also clear that the condition that $\gamma_x(t) \in {\cal E}$
reduces to $\left < N_i, \gamma_x'(t) \right > \leq 0$ and 
$\left < N_j, \gamma_x'(t) \right > \geq 0$ for some $i, j \in \{ 1,2,3\}$
which is computationally benign.

\section{Parametrization and Geometry of $C^E$ and $C^Z$}	\label{paramSec}

In this section we describe the parametrization and geometry of 
the faces and edges of $C$ generated by sharp features of $\partial M$ and 
the detection of singularities in these.   We extend the key constructs of 
{\em prism} and {\em funnel} proposed in~\cite{sweepComp} for smooth faces of $\partial M$ 
to the sharp features of $\partial M$. The funnel serves as the parameter space for faces 
of $C^E$ and guides further computation of the envelope.


Recall that a (smooth and non-degenerate) face is a map $S: D \rightarrow \mathbb{R}^3 $, where $D\subseteq \mathbb{R}^2 $ is a domain bounded by trim curves. 
A {\bf smooth parametric curve} in $\mathbb{R}^3$ is a (smooth and non-degenerate) map $e: d \to \mathbb{R}^3$ where 
$d=[s_0, s_1]$ is a closed interval of $\mathbb{R}$. Thus, specifying a face (resp. edge) requires us to specify the functions $S$ (resp. $e$)
and the domain $D$ (resp. $d$).

\subsection{Parametrization of faces $C^E$}		\label{parCESec}

Let $E$ be a sharp edge of $\partial M$ supported by two faces $F$ and $F'$.
Let $e$ be the curve underlying the sharp edge $E$ and $d$ be the subset of the parameter space of $e$ corresponding to 
$E$, i.e., $e(d) = E$.  
We extend the notion of {\em prism} proposed in~\cite{sweepComp} for smooth faces of $\partial M$ 
to the edge $E$.
At every point $e(s) \in E$, we may parametrize the cone of unit normals $N_{e(s)}$ as 
$N_{e(s)}(\alpha) =  \frac{ \alpha \cdot N_1 + (1 - \alpha) \cdot N_2 }{ \| \alpha \cdot N_1 + (1 - \alpha) \cdot N_2 \| }$ for 
$\alpha \in I_1=[0, 1]$, where, $N_1$ and $N_2$ are the unit outward normals to $F$ and $F'$ respectively at point $e(s)$.  
We  refer to the subset $d \times I_1 \times I$ of $\mathbb{R}^3$ as the prism of $E$. 
A point $(s, \alpha, t)$ in the prism corresponds to the normal $A(t) \cdot N_{e(s)}(\alpha) $ 
at the point $\gamma_{e(s)}(t)$ in the unit normal bundle ${\bf N_E}(t)$. 
Define the real-valued function $f^E$ on the prism of $E$ as $f^E(s, \alpha, t) = g(e(s), N_{e(s)}(\alpha), t)$. Clearly if $f^E (s,\alpha ,t)=0$ then $\gamma_{e(s)} (t)\in C^E $. This motivates us to define
the funnel as the {\em projection} of the zero-set of $f^E $ above to $d\times I$, as follows:

\begin{defn}	\label{funnelEDef}
For a sweep interval $I$ and a sharp edge $E \subset \partial M$, define 
${\cal F}^E = \{(s, t) \in d \times I | f^E(s, \alpha, t) = 0 \mbox{ for some } \alpha \in I_1 \}$. The set ${\cal F}^E$ is  
referred to as the {\bf funnel} for $E$. The set $\{ (s, t) \in {\cal F}^E | t = t' \}$ is referred to as the {\bf p-curve of contact} at $t'$ 
and denoted by ${\cal F}^E(t')$.
\end{defn}

The set ${\cal F}^E$ serves as the domain of parametrization for the faces $C^E$ generated by $E$. The parametrization function is given by
 $\sigma^E: {\cal F}^E \to \mathbb{R}^3$ as $\sigma^E(s, t) = A(t) \cdot e(s) + b(t)$. 

It now remains to compute the trim curves of ${\cal F}^E$. We now assume for simplicity the zero-set of $f^E $ is bounded by the boundaries of the prism $d\times I_1 \times I$. 
Thus the boundaries of ${\cal F}^E $ come from the equations $s=s_0 ,s_1 $ or $t=t_0, t_1 $ or finally $\alpha=0,1$. The first two conditions are easily implemented. The condition
$\alpha=0$ is equivalent to the assertion that $a^E_1 (s,t)=\langle A(t) \cdot N_1 (e(s)) , \gamma_{e(s)}'(t) \rangle =0$ , where $N_1 (e(s))$ is the normal to the face $F$ at the point $e(s)$. 
The function $a^E_1 (s,t)=0$ and the similarly defined  $a^E_2 (s,t)=0$ (for face $F'$) serve as the final trim curves.  
This collection of trim curves may yield several components, each corresponding to a unique face of $C^E $ on ${\cal E}$.
 
\eat{
Observe that $\sigma^E({\cal F}^E) = C^E$
and $\sigma^E(\partial {\cal F}^E) = \partial C^E$. On the boundary 
$\partial {\cal F}^E$, either $s \in \{ s_0, s_1 \}$ or $\alpha \in \{ 0, 1 \}$ 
or $t \in \{ t_0, t_1 \}$.  

The portion of the boundary where $s \in \{ s_0, s_1 \}$ 
corresponds to the edges generated by sharp vertices which bound $E$.  
The portion where $t \in \{ t_0, t_1 \}$ corresponds to the curves of contact 
at initial and final times and $\alpha \in \{ 0, 1\}$ corresponds to the edges 
$\Pi_C \left ({\bf C^{(N_E \cap N_F)}} \right )$ and $\Pi_C \left ( {\bf C^{(N_E \cap N_{F'})}} \right )$ 
where $F$ and $F'$ are the smooth faces which are adjacent to $E$ in $\partial M$.
}

\begin{figure}
 \centering
\includegraphics[width=1.0\linewidth]{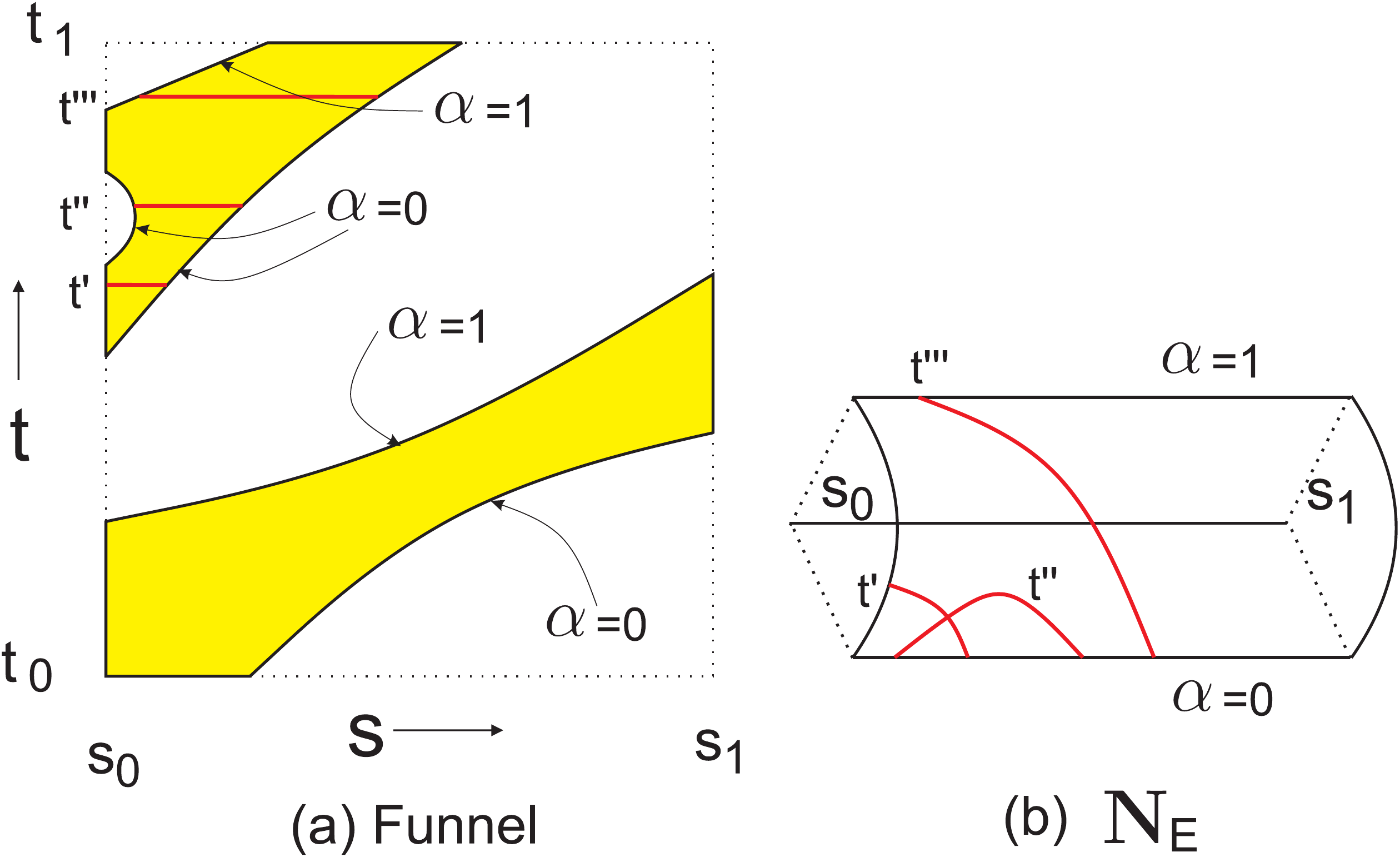}
 \caption{(a) The funnel ${\cal F}^E$ is shaded in yellow. The p-curves of contact at $t', t''$ and $t'''$ are shown in red. 
(b) The curves ${\pmb \pi}({\bf C}(t')), {\pmb \pi}({\bf C}(t''))$ and ${\pmb \pi}({\bf C}(t'''))$  are shown on ${\bf N_E}$. }
 \label{funnelFig}
\end{figure}

Fig.~\ref{funnelFig}(a) illustrates the funnel ${\cal F}^E$ shaded in yellow and p-curves of contact ${\cal F}^E(t'), {\cal F}^E(t'')$ and 
${\cal F}^E(t''')$ shown in red.  In this example, ${\cal F}^E$ has two connected components. The curves $\sigma^E ({\cal F}^E (t))$ are parts of the curve of contact on ${\cal E}$ at time $t$. 
 In Fig.~\ref{funnelFig}(b), 
the normals of contact, i.e., $A(t) \cdot e'(s) \times \gamma_x'(t)$ at times $t', t''$ and $t'''$ are shown projected on the unit normal bundle ${\bf N_E}$.

\subsection{Singularities in $C^E$}	\label{singSec}

A parametric surface $S$ is said to have a singularity at a point $S(u_0,v_0)$ if $S$ fails to be an immersion at $(u_0, v_0)$, i.e., the rank 
of the Jacobian $J_S$ falls below 2.

\begin{lem}		\label{singLem}
Let $p = (s', t') \in {\cal F}^E$. A face of $C^E$ has a singularity at point $\sigma^E(p)$ if and only if the velocity $\gamma_{e(s')}'(t')$ is tangent to 
the edge $E$ at the point $\sigma^E(p)$, i.e., $\gamma_{e(s')}'(t')$ and $A(t') \cdot \frac{de}{ds}(s')$ are linearly dependent.
\end{lem}

\begin{figure}
 \centering
\includegraphics[width=1.0\linewidth]{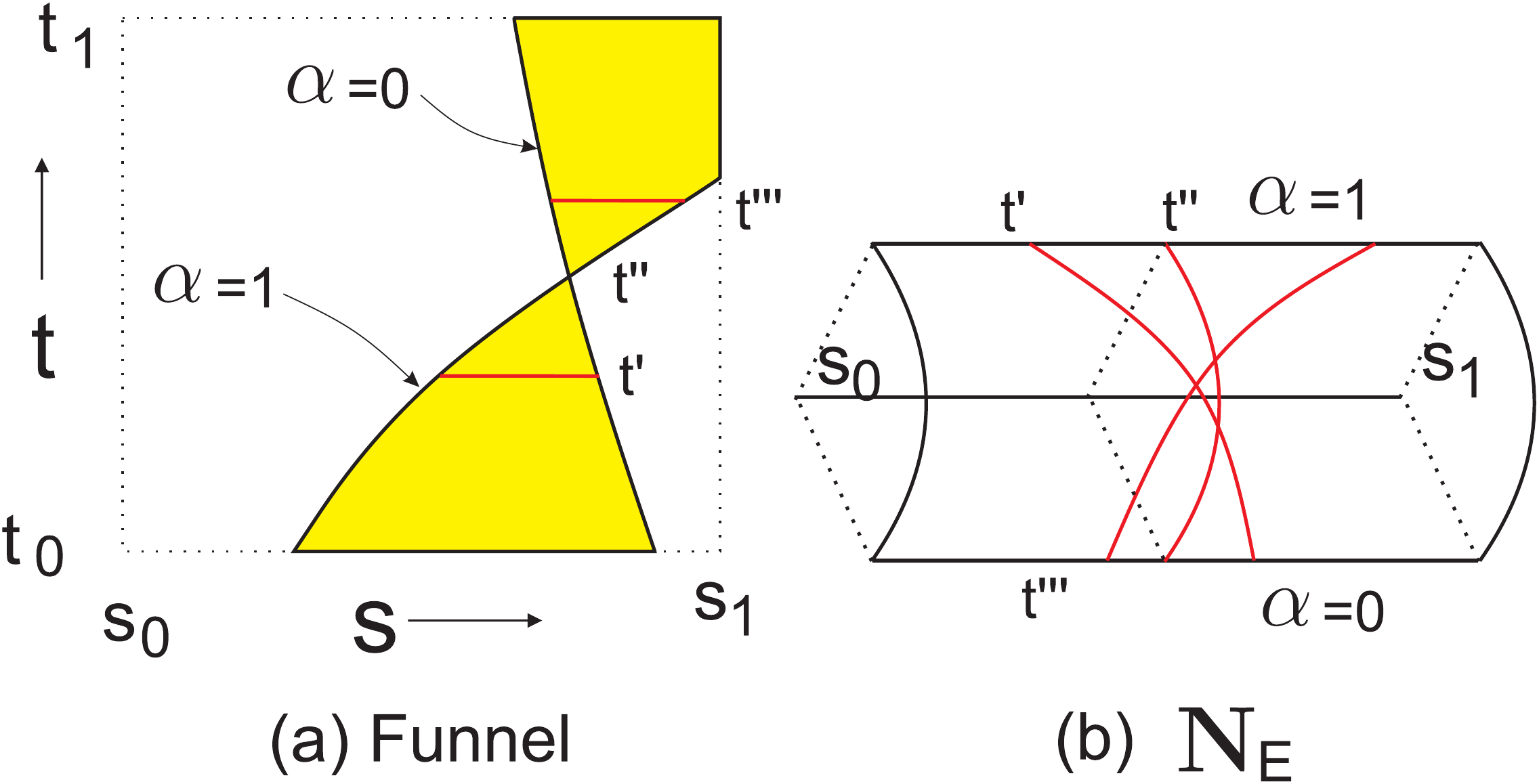}
 \caption{Singularity in $C^E$. (a) The funnel is shaded in yellow. The p-curves of contact 
${\cal F}^E(t'), {\cal F}^E(t'')$ and ${\cal F}^E(t''')$ are shown in red.  (b) The curves 
${\pmb \pi}({\bf C}(t')), {\pmb \pi}({\bf C}(t''))$ and ${\pmb \pi}({\bf C}(t'''))$  are shown on ${\bf N_E}$.  }
 \label{singFig}
\end{figure}

\begin{figure}
 \centering
\includegraphics[scale=0.35]{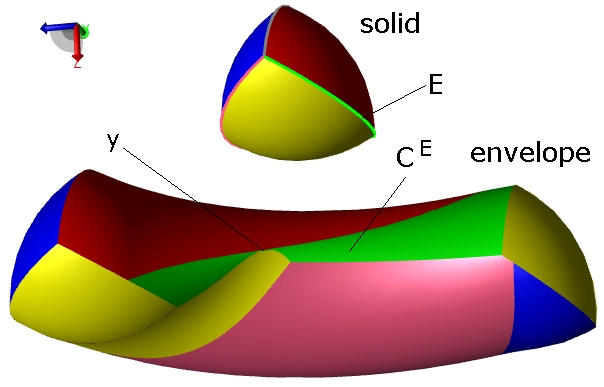}
 \caption{The contact set $C^E$ has a singularity at point $y$.}
 \label{singExFig}
\end{figure}

Fig.~\ref{singFig} illustrates schematically a funnel ${\cal F}^E$ having a singularity at $t''$. 
A sweep example with singularity is shown in Fig.~\ref{singExFig}.

\subsection{Parametrization of edges $C^Z$}		\label{parCZSec}

\begin{figure}
 \centering
 \includegraphics[scale=0.45]{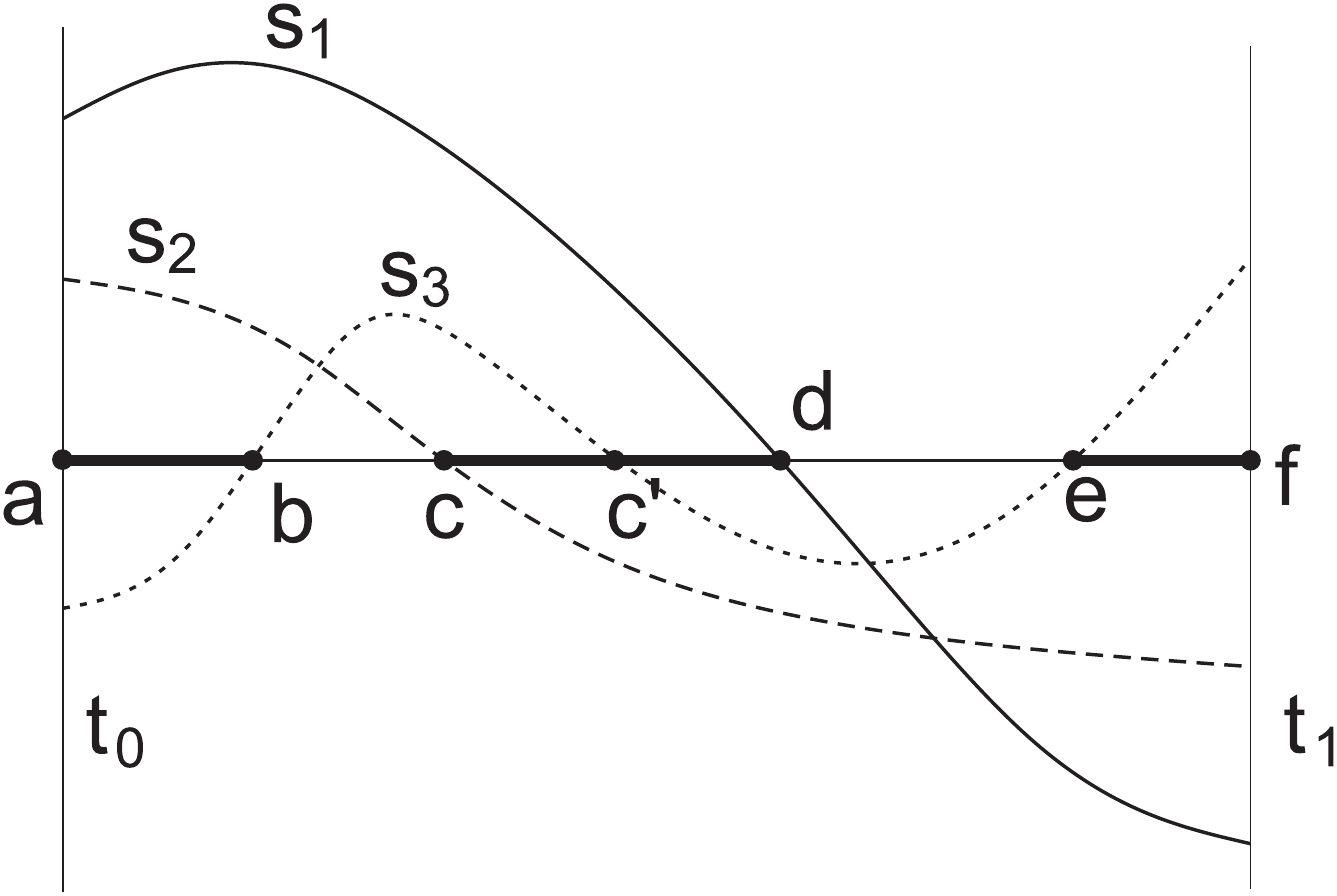}
 \caption{The functions $s_1$, $s_2$ and $s_3$ are plotted against time.}
 \label{threeSFig}
\end{figure}

Let $Z$ be a sharp vertex lying in the intersection of faces $F_1, F_2$ and $F_3$ and let 
$N_1, N_2$ and $N_3$ be the unit outward normals to $F_1, F_2$ and $F_3$ respectively at $Z$. 
As noted in Section~\ref{calculusSec}, the point $\gamma_Z(t)$ belongs to the contact set 
$C^Z$ if and only if $\left < A(t) \cdot N_i, \gamma_Z'(t) \right > \leq 0$ and 
$\left < A(t) \cdot N_j, \gamma_Z'(t) \right > \geq 0$ for some $i, j \in \{ 1,2,3\}$. 
Define functions $s_i: I \to \mathbb{R}$ as $s_i(t) = \left < A(t) \cdot N_i , \gamma_Z'(t) \right >$ 
for $i = 1,2,3$. 
Clearly, the contact set $C^Z$ corresponds to the set of closed sub-intervals of 
the sweep interval $I$ where any two of the functions $s_i$ differ in sign.  This 
is illustrated schematically in Fig.~\ref{threeSFig}.  At the end-points of these sub-intervals, 
either $t \in \{ t_0, t_1 \}$ (illustrated by points $a$ and $f$ in Fig.~\ref{threeSFig}) 
or one of the functions $s_i$ is zero (illustrated by points $b, c, c',d$ and $e$ in Fig.~\ref{threeSFig}).
Thus the collection of sub-intervals $d_Z$ of $I$ is easily computed. The parametrization function of course is 
$\gamma_Z :d_Z \rightarrow \mathbb{R}^3$ given by the trajectory of the point $Z$ under $h$. This finishes the parametrization of $C^Z $.  

\eat{
For each open interval, say $(c,c')$, 
The normals of contact at time $t = b$ in Fig.~\ref{threeSFig}, i.e., ${\pmb \pi}({\bf C}^Z(t))$ is shown in Fig.~\ref{adjConfigFig}(a). 
At time $t= c'$ in Fig.~\ref{threeSFig}, ${\pmb \pi}({\bf C}^Z(t))$ is shown in Fig.~\ref{adjConfigFig}(c).  For $t$ in interval $(c, c')$, 
${\pmb \pi}({\bf C}^Z(t))$ is shown in Fig.~\ref{adjConfigFig}(b).
}

\section{Adjacencies and topology of $C$}		\label{topoSec}

We now focus on the matching of co-edges for each face of $C$. We already know that faces of $C$ come from (i) $C^E$ when $E$ is a sharp edge, or (ii) $C^F $ when $F$ is a smooth face. 
Similarly edges in $C$ come from (i) edges bounding faces of $C^E ,C^F $ and (ii) edges coming from $C^Z$, where $Z$ is a sharp vertex. The matching of co-edges is eased by the following {\bf proximity lemma}.  
While the global brep structure of $C$ may be very different from that of 
$\partial M$, we show that locally they are similar.


Recall the natural correspondence ${\pi}: { C} \to {{\partial M}}$ 
from Section~\ref{mathStrSec}.  
We show that the adjacency 
relations between geometric entities of $C$ are preserved by the correspondence $\pi$.  

\begin{lem}	\label{contProp}
The correspondence map $\pi:  C \to \partial M$ is continuous.
\end{lem}
\noindent {\em Proof.} For a face 
$F \subseteq M$, we denote the restriction of the map ${\pi}$ to ${C^{F}}$ by $\pi^F$, i.e.,  
$\pi^F: C^F \to F$,  $\pi^F(y) = \pi(y)$.
The restriction of ${\pi}$ to ${C^{E}}$  for a sharp edge $E \subset \partial M$ is defined similarly. Consider first the restriction $\pi^E$ of $\pi$ to 
$C^E$.  Recall the 
parametrization of $C^E$ via the funnel ${\cal F}^E$ and $\sigma^E$ from Section~\ref{parCESec}. 
Let $y \in C^E$ and $p = (s',t') \in {\cal F}^E$ such that $\sigma^E(p) = y$.
The map $\sigma^E$ being continuous,  in order to show that $\pi^E$ is continuous at $y$, 
it is sufficient to show that the composite map $\pi^E \circ \sigma^E  : {\cal F}^E \to E$ 
given by $\pi^E \circ \sigma^E (s, t) = e(s)$ is continuous at $p$, where, $e$ is the parametric curve 
underlying edge $E$.  This follows from the continuity of $e$.  


The continuity of the restriction $\pi^F$ to $C^F$ for a face $F \subseteq \partial M$ can be 
similarly proved, by choosing a pair of local coordinates at any point $p \in {\cal F}^F$.

The continuity of the map $\pi$ follows from the fact that $\pi$ is obtained by gluing the maps 
$\{ \pi^F | F \subseteq M \} \cup \{ \pi^E | E \mbox{ is a sharp edge in } \partial M \}$ each of 
which is continuous.
\hfill $\square$

We conclude the following theorem from the above proposition.

\begin{thm}		\label{adjThm}
For any two geometric entities $O$ and $O'$ of $\partial M$, if $C^O$ and $C^{O'}$ are adjacent 
in $C$, then $O$ and $O'$ are adjacent in $\partial M$.
\end{thm}

In other words, for a face $F \subset \partial M$ and a sharp edge $E \subset \partial M$, 
if faces $C^F_i$ and $C^E_j$ are adjacent in $C$, then $F$ and $E$ are adjacent in
$\partial M$.  For a sharp vertex $Z \subset \partial M$, if an edge $C^Z_k$ bounds a 
face $C^E_j$ in $C$ then the vertex $Z$ bounds the edge $E$ in $\partial M$. 
 
This aids the computation of adjacency relations amongst entities of $C$ and is illustrated 
by the sweep example shown in Figures~\ref{contactSetFig},~\ref{coneCubeHelixFig},~\ref{singExFig}  
and~\ref{showCaseFig} by color coding.  The entities $O$ and $C^O$ 
are shown in same color.  
\eat{In Section~\ref{sharpAdjSec} we further refine the computation of adjacency 
relations amongst faces adjacent to edge $C^Z$ generated by a sharp vertex $Z$.
}

\subsection{Co-edges bounding faces $C^E$}	\label{coedgSec}

\begin{figure}
 \centering
 \includegraphics[scale=0.35]{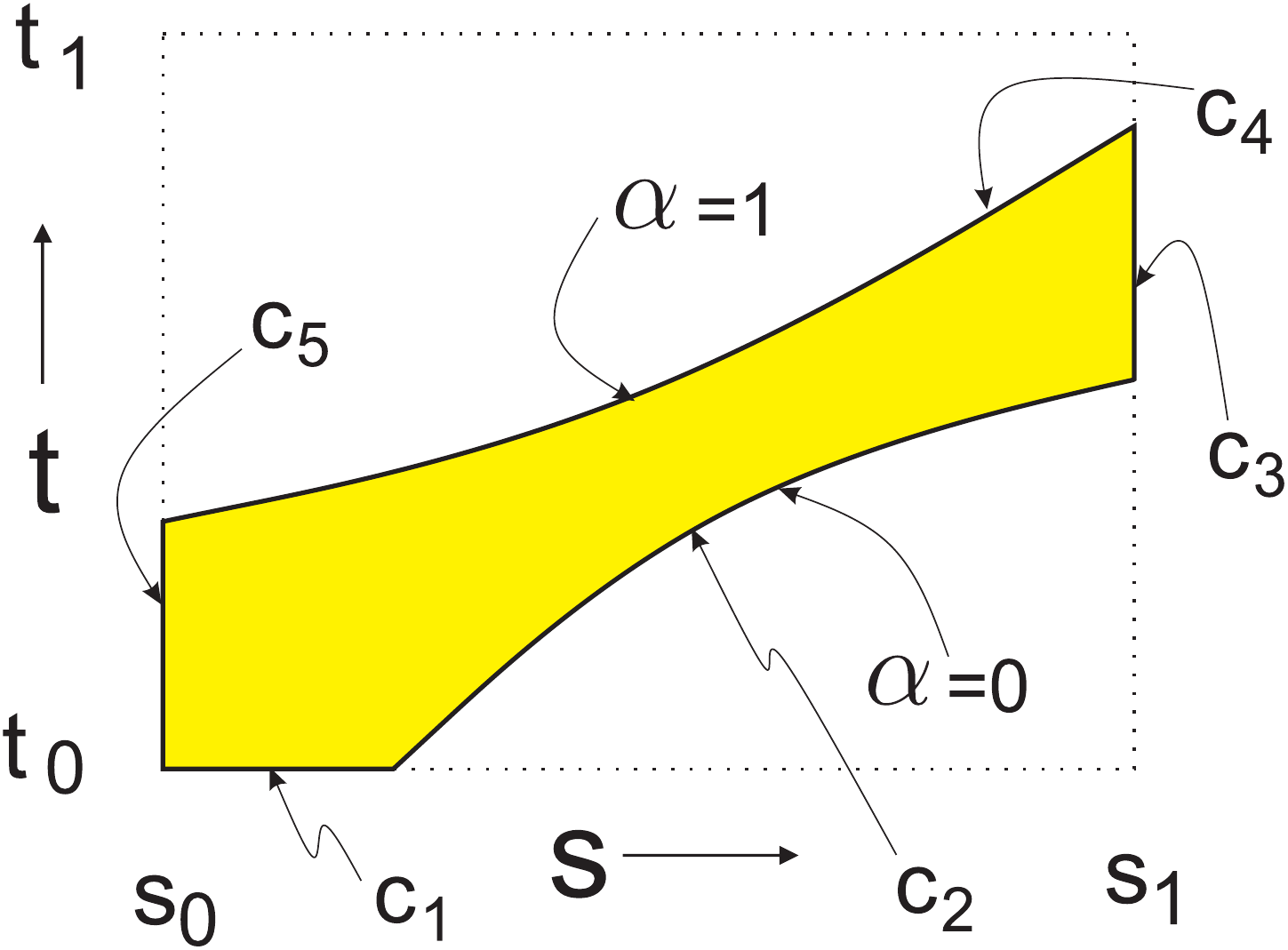}
 \caption{Co-edges bounding face $C^E$. }
 \label{coedgesFig}
\end{figure}

Consider a  sharp edge $E$ supported by smooth faces $F$ and $F'$ in $\partial M$. We pick a face of $C^E $ given by the component of 
${\cal F}^E $ shown in Fig.~\ref{coedgesFig}. The co-edges $c_5 ,c_3 $ come from the equations $s=s_0 $ and $s=s_1 $ respectively. These must correspond 
to edges swept by sharp vertices bounding the edge $E$. The co-edge $c_1 $ comes from the condition $t=t_0$ and thus comes from curve of contact at the 
initial time instant and thus, the left cap. Finally, the curves $c_2, c_4 $ correspond to $a^E_1 (s,t)=0$ and $a^E_2 (s,t)=0$ which come from the normals of 
contact matching that of the supporting smooth faces as described in Section~\ref{paramSec}. Thus these co-edges must match those coming from the boundaries of $C^F $ and $C^{F'}$.  

\subsection{Co-edges matching edges of $C^Z $}	\label{coedgZ}

\begin{figure}
 \centering
 \includegraphics[width=1.0\linewidth]{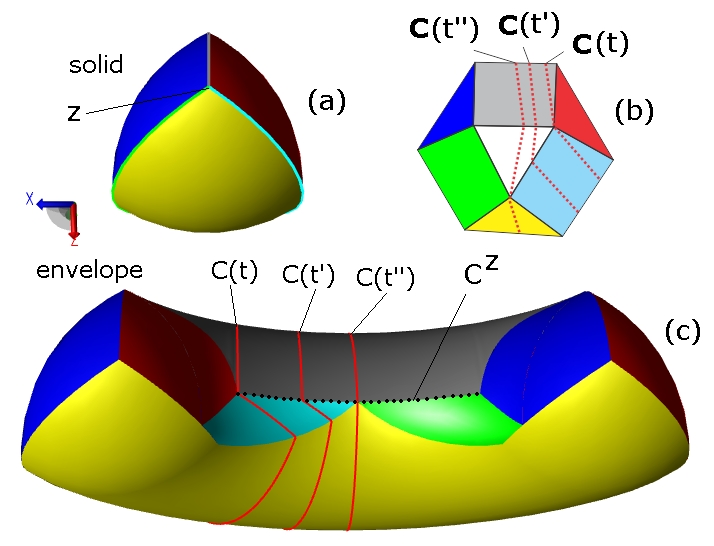}
 \caption{Adjacency relations between faces of $C$. (a) Solid being swept. (b) Normals of contact ${\pmb \pi}({\bf C}(t))$, 
${\pmb \pi}({\bf C}(t'))$ and ${\pmb \pi}({\bf C}(t''))$ are shown on the unit normal bundle ${\bf N_{\partial M}}$.
(c) Curves of contact $C(t)$, $C(t')$ and $C(t'')$ are shown in red.  The edge $C^Z$ generated by the sharp vertex $Z \subset \partial M$ 
is shown as a dotted curve in black on $C$. }
 \label{triSphArcConeFig}
\end{figure}

We next come to the co-edges matching with edges arising from $C^Z$. As in Fig.~\ref{threeSFig}, the edges of $C^Z $ are parametrized by intervals $d_1 ,\ldots, d_k $. 
Each interval $d_i $ has two of the three functions $s_ 1, s_2$ and $s_3 $ of one sign and the third of the opposite sign. For example, if we take the interval $(c,c')$, we
see that $s_1,s_3 >0$ and $s_2 <0$. For $t\in (c,c')$, if we look at  the zero locus of the function $\langle A(t)\cdot n, \gamma_Z'(t) \rangle$, on $N_Z $, then there must be an $n_1 \in 
cone(N_1 ,N_2 )\subset N_Z $ such that $\langle A(t)\cdot n_1 ,\gamma_Z'(t)\rangle =0$ and there must be an $n_2 \in cone(N_2,N_3 )\subset N_Z $ such that $\langle A(t)\cdot n_2 ,\gamma_Z'(t) \rangle =0$.      
This leads us to the sharp edge $E_1 $ with normals $N_1,N_2 $ at the vertex $Z\in E_1 $, and to the sharp edge $E_2 $ with normals $N_3 , N_2 $ at $Z \in E_2$ and the conclusion that that faces of $C^{E_1 }$ 
and $C^{E_2}$ meet at the edge $[c,c']$ of $C^Z$. See for example, the curve of contact $C(t')$ in Fig.~\ref{triSphArcConeFig}. 
A similar conclusion for the interval $(c',d)$ tells us that faces of $C^{E_1 }$ 
and $C^{E_3 }$ 
meet on the edge $[c',d]$ of $C^Z $. The curious point is the time instant $c'$ where the smooth face $F_3 $ with normal $N_3 $ also meets the edge $C^Z$. This is illustrated by curve of contact $C(t'')$ 
in Fig.~\ref{triSphArcConeFig} where there are four incident faces.

\section{Computation of the brep of $C$}	\label{brepSec}

In this section we explain Steps 14 to 27 of Algorithm~\ref{frameworkAlgo} for 
generating the entities on the envelope corresponding to sharp edges and vertices 
of $\partial M$.  Algorithm~\ref{frameworkAlgo} marches over each entity $O$ of 
$\partial M$ and computes the corresponding entity $C^O$ of $C$. 
The computation of $C^O$ follows the computation of its boundary $\partial C^O$.
For further discussion fix a sharp edge $E$ of $\partial M$ (cf Step 14 of Algorithm~\ref{frameworkAlgo}).

\eat{
\subsection{Computing faces $C^E$}

In this section we describe the computation and orientation of the 
faces $C^E$ generated by the sharp edge $E \subset \partial M$. 
Recall from Section~\ref{parCESec} that the faces $C^E$ are parametrized 
via the funnel ${\cal F}^E$.
Before computing $C^E$, its boundary is computed and oriented. 
}

\subsection{Computing and orienting co-edges $C^Z$}

Consider a sharp vertex $Z \subset \partial E$.  Recall from Section~\ref{parCZSec} 
that computing the edges $C^Z$ is equivalent to computing the collection of closed 
subintervals of the sweep interval $I$ in which the functions $s_i$ differ in sign.
We use Newton-Raphson solvers for computing the end-points of these subintervals.
Of course, these end-points give rise to vertices which bound edges of $C^Z$. 
This is performed in Step 16 of Algorithm~\ref{frameworkAlgo}.

\begin{figure}
 \centering
\includegraphics[scale=0.5]{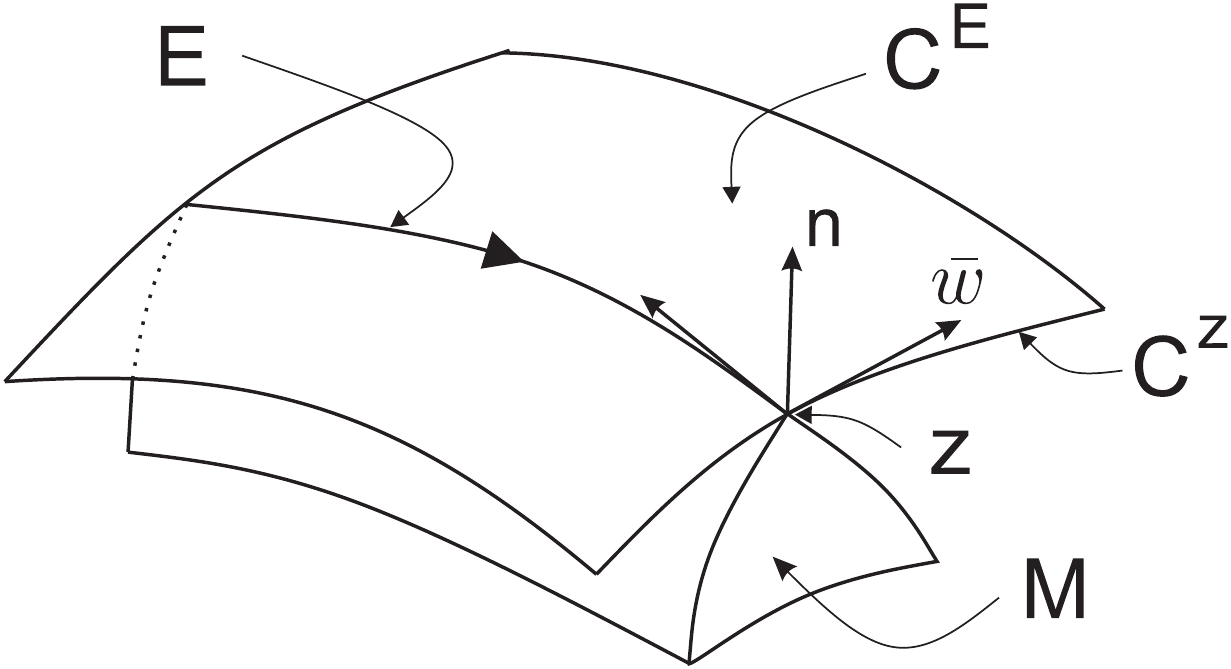}
 \caption{Orienting co-edges $C^Z$.  In this case $e(s_1) = Z$ and  $ -e'(s)|_{s_1}$ points in the interior of face $C^E$.}
 \label{orientCZFig}
\end{figure}

Each co-edge $C^Z_i$ bounding the face $C^E_j$ must be oriented so that $C^E_j$ is on its 
left side with respect to the outward normal in a right-handed co-ordinate system. 
Let $y = \gamma_Z(t) \in C^Z_i$ and $\bar{w} \in \mathbb{R}^3$ be tangent to $C^Z_i$ at $y$.  
Let $n$ be the outward unit normal to $C^E_j$ at $y$ (cf Section~\ref{orientCESec}).
Assume without loss of generality that $A(t) = I$ and $b(t) = 0$.
Let $e$ be the parametric curve underlying $E$ so that $e(d) = E$ where $d = [s_0, s_1]$. 
Consider two cases as follows.
\begin{enumerate}
\item If $Z = e(s_0)$, then $e'(s_0)$ points in the interior of the face $C^E_j$, where, $e'$ denotes 
the derivative of $e$.
If $\left < e'(s_0) , n \times \bar{w} \right > > 0$ then $\bar{w}$ is the orientation of $C^Z_i$ 
else $-\bar{w}$ is the orientation.

\item If $Z = e(s_1)$ then $-e'(s_1)$ points in the interior of $C^E_j$.  
If $\left < -e'(s_1) , n \times \bar{w} \right > > 0$ then $\bar{w}$ is the orientation of $C^Z_i$ 
else $-\bar{w}$ is the orientation. This is illustrated schematically in Fig.~\ref{orientCZFig}.
\end{enumerate}

The co-edges $C^Z$ are oriented in Step 17 of Algorithm~\ref{frameworkAlgo}.

\subsection{Computing and orienting co-edges  $C^E \cap C^F$ and $C^E \cap C^{F'}$} 

For the sharp edge $E$ supported by smooth faces $F$ and $F'$ in $\partial M$, the co-edges 
 $C^E \cap C^F$ and $C^E \cap C^{F'}$
bounding a face of $C^E$ correspond to the iso-$\alpha$ curves for $\alpha \in \{0,1\}$ of $C^E$ as discussed in Section~\ref{coedgSec}. 
The orientation of these co-edges for $C^E$  
is opposite to that of the partner co-edges for $C^F$ and $C^{F'}$.  The co-edges bounding $C^F$ and $C^{F'}$ 
are computed and oriented in Steps 6 and 7 of Algorithm~\ref{frameworkAlgo}.  Their partner co-edges bounding faces $C^E$ are 
computed and oriented in Step 20 and 21 of  Algorithm~\ref{frameworkAlgo}.

\subsection{Computing loops bounding faces $C^E$}

A loop is a closed, connected sequence of oriented co-edges which bound a face. 
As noted in Section~\ref{coedgSec}, the co-edges bounding faces of $C^E$ are either iso-$\alpha$ 
curves for $\alpha \in \{0, 1\}$, or iso-$s$ curves for $s \in \{ s_0, s_1\}$ or iso-$t$ curves 
for $t \in \{t_0, t_1\}$.  In order to compute the loop bounding a face $C^E_i$, we start with 
a co-edge bounding $C^E_i$ and find the next co-edge in sequence.  For instance, if this 
co-edge is iso-$\alpha$ curve for $\alpha = 0$ and its end-point is $(\alpha, s) = (0, s_1)$ then 
the next co-edge in sequence is iso-$s$ curve with $s=s_1$. This is repeated till the loop 
is closed. Fig.~\ref{coedgesFig} illustrates this schematically.  This computation is performed 
in Step 23 of Algorithm~\ref{frameworkAlgo}.

\subsection{Computing and orienting faces $C^E$}	\label{orientCESec}

The parametrization of faces $C^E$ was discussed in Section~\ref{parCESec} 
via the funnel ${\cal F}^E$. This is done in Step 24 of Algorithm~\ref{frameworkAlgo}.
Each face in the brep format is oriented so that the unit normal to the face points in 
the exterior of the solid.  Consider a point $y = \gamma_z(t) \in C^E$ and assume 
without loss of generality that $A(t) = I$ and $b(t) = 0$.
Recall from Section~\ref{coneSec} and Section~\ref{parCESec} that 
if $\bar{w}$ is tangent to $E$ at $z$, then $n := A(t) \cdot \bar{w} \times \gamma_z'(t)$ 
is normal to $C^E$.  Further, either $n \in A(t) \cdot N_z$ or $-n \in A(t) \cdot N_z$.
Since the interior of the swept volume is ${\cal V}^o = \cup_{t \in I}M(t)^o$, 
the outward normal to $C^E$ at $y$ is $n$ if $n \in A(t) \cdot N_z$ else it is 
$-n$.  This is performed in Step 25 of Algorithm~\ref{frameworkAlgo}.

\begin{figure}
 \centering
\includegraphics[scale=0.25]{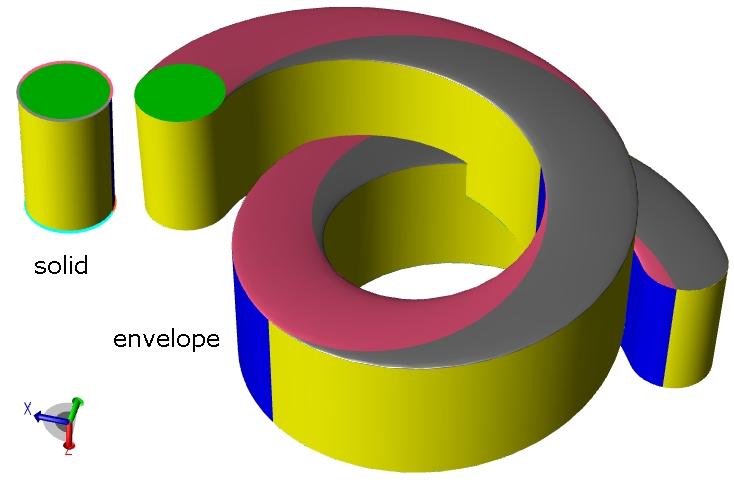}
 \caption{An example illustrating global self-intersection.}
 \label{toolFig}
\end{figure}

\eat{
\section{Discussion}	\label{discussSec}

Recall from Section~\ref{lsiSec} that the faces 
$C^E$ will be free of local self-intersections.  
Global self-intersections (see~\cite{trimming, selfIntersections}) in $C$ may be resolved by surface-surface intersections, which is a 
standard routine in modern CAD kernels~\cite{acis}.  A sweep example with global self-intersection 
appears in Fig.~\ref{toolFig}. 
A concave sharp edge of $\partial M$ does not generate any faces on the envelope, 
however it will lead to self-intersections which are beyond the scope of this paper.
}

Our framework is tested on over 50 different solids with number of sharp edges and smooth faces between 4 and 25,  swept along 
complex trajectories.  A pilot implementation using the ACIS~\cite{acis} kernel took between 30 seconds to 2 minutes 
on a Dual Core 1.8 GHz machine for these examples, some of which appear in Fig.~\ref{showCaseFig}. Many more examples are 
included in the supplementary file.

\begin{figure*}
  \begin{subfigure}[b]{0.33\linewidth}
    \includegraphics[width=\linewidth]{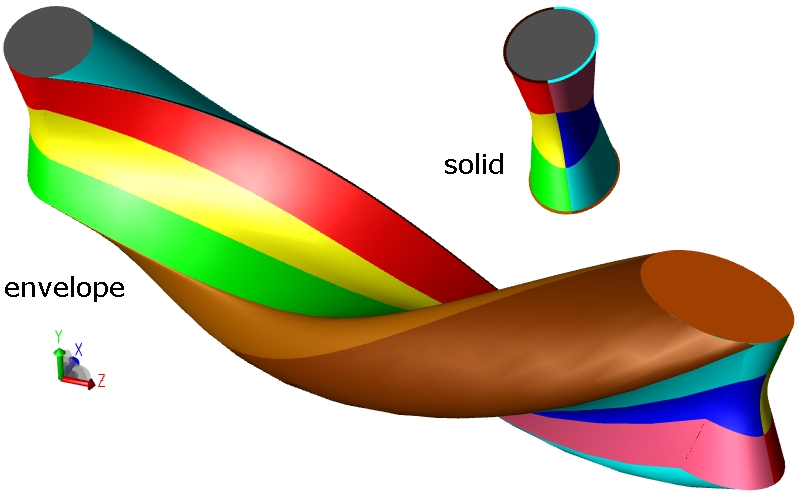} 
  \end{subfigure} 
  \begin{subfigure}[b]{0.33\linewidth}
    \includegraphics[width=\linewidth]{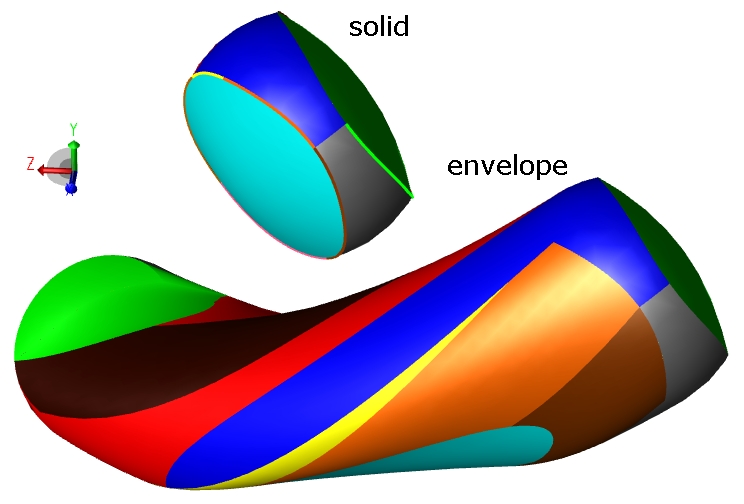}  
  \end{subfigure}
\begin{subfigure}[b]{0.33\linewidth}
    \includegraphics[width=\linewidth]{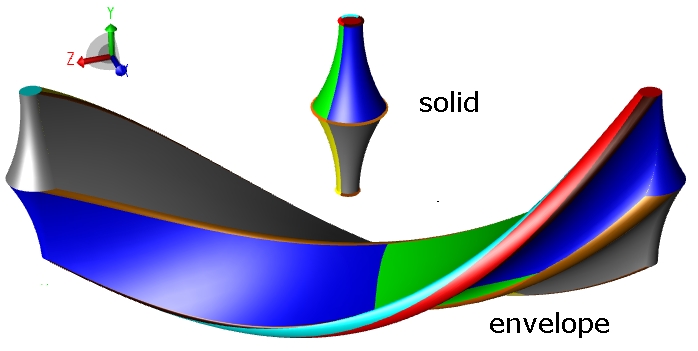} 
  \end{subfigure}

\begin{subfigure}[b]{0.33\linewidth}
    \includegraphics[width=\linewidth]{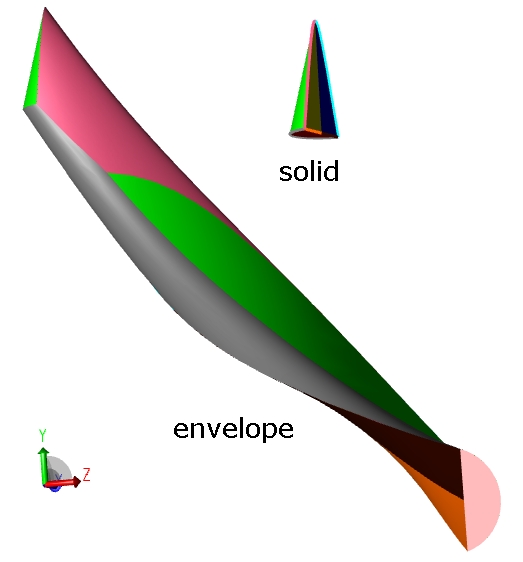}
  \end{subfigure} 
  \begin{subfigure}[b]{0.33\linewidth}
    \includegraphics[width=\linewidth]{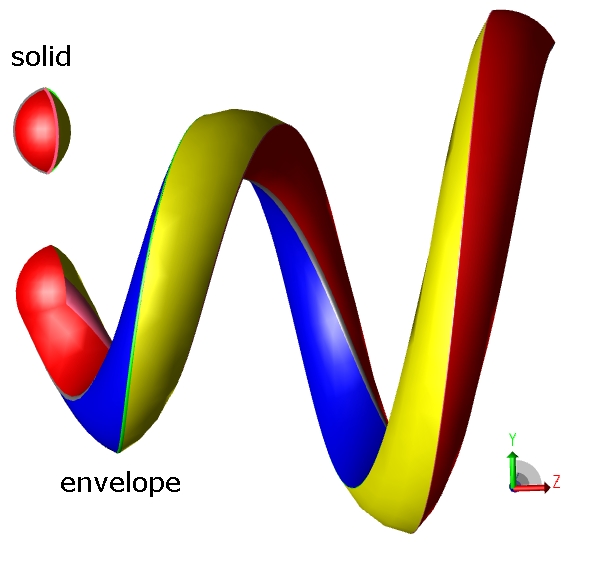}
  \end{subfigure}
\begin{subfigure}[b]{0.33\linewidth}
    \includegraphics[width=\linewidth]{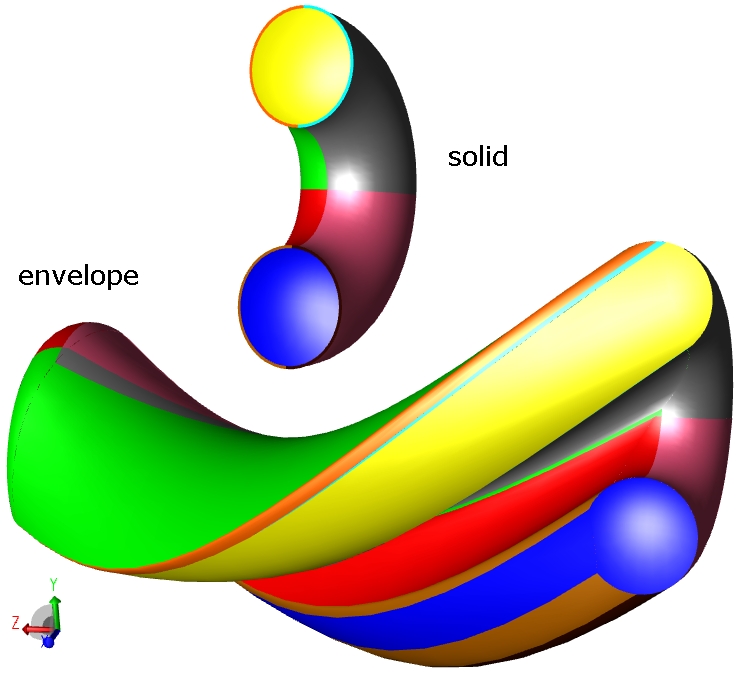}
  \end{subfigure}


\begin{subfigure}[b]{0.33\linewidth}
    \includegraphics[width=\linewidth]{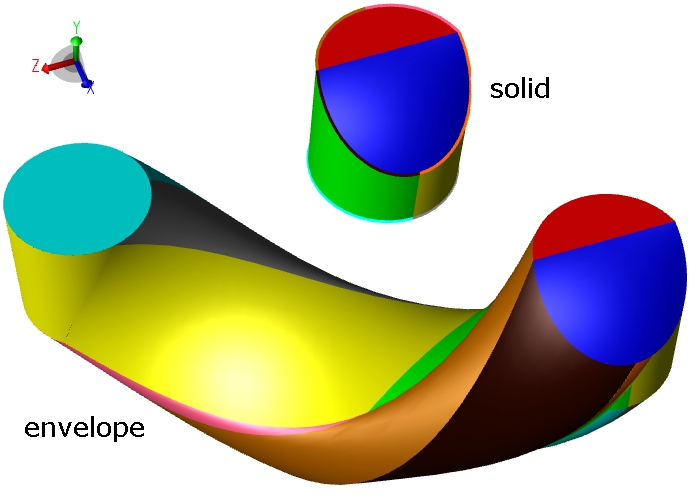}
  \end{subfigure} 
  \begin{subfigure}[b]{0.33\linewidth}
    \includegraphics[width=\linewidth]{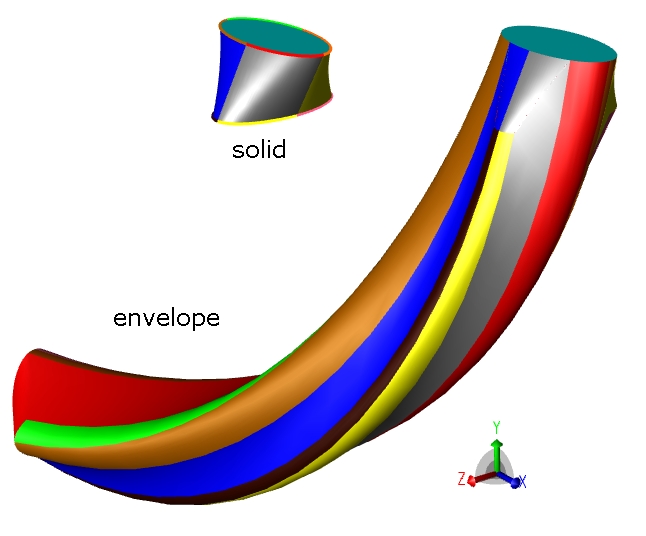}
  \end{subfigure}
\begin{subfigure}[b]{0.33\linewidth}
    \includegraphics[width=\linewidth]{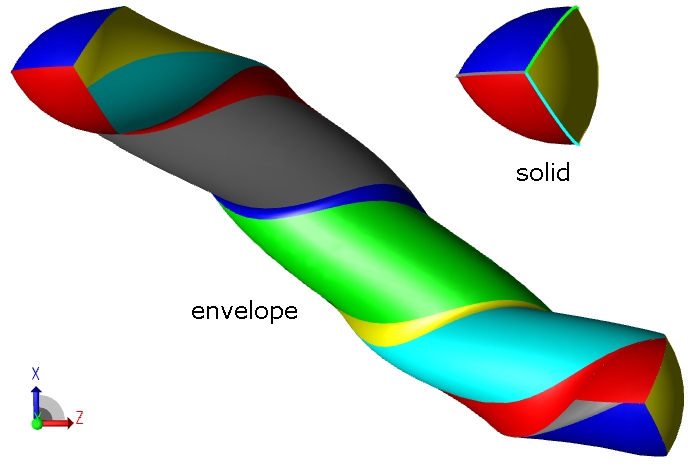}
  \end{subfigure}
\caption{Examples of solid sweep}
\label{showCaseFig}
\end{figure*}

\section{Extension to non-simple sweeps} \label{extensions}
In this section, we discuss an extension of the above framework to `non-simple' sweeps. Recall that, in a non-simple
sweep, the correct construction of the envelope proceeds with an appropriate trimming of the contact set.
This calls for local and global self-intersections of the contact set (see~\cite{trimming, selfIntersections,localGlobal} for definitions). 
Global self-intersections may be resolved by surface-surface intersections, which is a 
standard routine in modern CAD kernels.  A sweep example with global self-intersection 
appears in Fig.~\ref{toolFig}. Local self-intersections are more subtle. Roughly speaking, in a local self-intersection, a point
on the contact set is occluded by an infinitesimally close point.

In~\cite{localGlobal}, the authors assume that the input solid is smooth and construct an {\em invariant} function $\theta$ on the 
contact set which efficiently separates global self-intersections from local self-intersections. The function $\theta$ is intimately related to
local curvatures and the inverse trajectory (see~\cite{trimming, classifyPoints}) used in earlier works. Further, it has been shown there that $\theta$ is robust and provides
the key `seed` information to resolve local self-intersections via surface-surface intersections. Much of this work also extends to
sharp solids albeit restricted to only the part of the contact set which is generated by smooth features. Clearly, it is important
to understand the self-intersections on the contact set generated by sharp features. We next show that the sharp features never
give rise of local self-intersections!

\begin{defn} \label{invTrajDef}
Given a trajectory $h$, the {\bf inverse trajectory} $\bar{h}$ is defined as the map $\bar{h}:I  \to (SO(3), \mathbb{R}^3)$ given by $\bar{h}(t) = (A^t(t), -A^t(t) \cdot b(t))$.  
Thus, for a fixed point $x \in \mathbb{R}^3$, the inverse trajectory of $x$ is the map $\bar{\gamma}_x:I \to \mathbb{R}^3$ 
given by $\bar{\gamma}_x(t) = A^t(t) \cdot (x - b(t))$. Observe that, under the trajectory $h$, 
the point $\bar{\gamma}_x(t)$ transforms to $x$ at time t.
\end{defn}

The contact set $C$ is said to have a {\bf local self-intersection} (L.S.I.) (see~\cite{trimming, selfIntersections}) 
at a point $y =\gamma_x(t')$ if for all $\delta t > 0$, there exists $t'' \in (t' - \delta t, t' + \delta t)$, such that 
$\bar{\gamma}_x(t'') \in M^o(t')$,  
where $M^o$ denotes the interior of $M$. Thus, $y$ is occluded by an infinitesimally close point in the interior of the solid $M$. 

\begin{prop}	\label{lsiProp}
For a sharp convex point $x$ on the edge $E$ of $\partial M$, each point $y = \gamma_x(t')$ lying in the interior of a face of $C^E$ is free of L.S.I.
\end{prop}
Refer to Appendix~\ref{proof2Sec} for proof.

As there is no outward normal at a concave sharp point, it is easily seen that, in the generic situation, the concave features 
do not generate any point on the envelope. In fact, the concave features will almost always lead to global self-intersections 
of the contact set and hence result into non-simple sweeps! This provides the justification of our standing assumption that the
input solid does not have a sharp concave edge.

The implementation of simple sweeps is complete and uses the ACIS kernel.
The extension to non-simple sweeps is in progress and will require (i) scheduling of surface-surface intersections and (ii) integration of $\theta$. 
 ACIS already provides standard robust and computationally efficient API's  for transversal surface-surface intersections. 


\section{Conclusion} \label{concludeSec}
This paper extends the framework of~\cite{sweepComp}  for the construction of free-form
sweeps from smooth
solids to solids with sharp features. This was done by developing a
calculus of normal cones and their interaction with a one-parameter family
of motions. Furthermore, this calculus leads to a neat extension of the
key devices of the {\em prism, funnel} and results in a computationally
clean and efficient computation of the trim curves
and also of the curves arising from sharp vertices. This in turn leads us
to a robust implementation of the general sweep. Numerous models
have been successfully generated using this implementation. We have
also discussed an extension of the above framework to allow for local and global self-intersections.

The normal bundle indicates a connection between the sweep and the
off-set. It is likely that these operations commute, as is
indicated by the calculus of cones presented here. Perhaps, this
mathematical observation will lead to a better implementation in the
future.
Finally, the above framework actually constructs the normal bundle of the
sweep and that this has several interesting features. For example, it has
no sharp
vertices (other than those coming from the left or right caps) even though
$M$ may have. The sharp vertices of $M$ however lead to degenerate
vertices in ${\cal E}$.

Another point is the so-called procedural framework and the construction of the seed or
approximate surfaces which are used to initialize the
evaluators. The construction of these need substantial care and a complete
discussion of this is deferred to a later paper.

\eat{
the mathematical framework of~\cite{sweepComp} to accomplish much of this, however, some
deeper Morse-theoretic limitations creep up. These
arise from the assumption that the curve of contact is simple. 

For solids
with concavities, this is almost always not true
and the construction of the the global self-intersection is fairly
non-trivial and needs to be explored.}

\appendix

\section{Proof of Proposition~\ref{gProp}}	\label{proof1Sec}
Define the following subsets of $\mathbb{R}^4$ where the fourth dimension is time.
Let $Z:= \{ (A(t) \cdot x + b(t), t) | x \in M \text{ and } t \in I \}$ and $X :=  \{ (A(t) \cdot x + b(t), t) | x \in \partial M \text{ and } t \in I \}$. 
Note that $Z$ is a four dimensional topological manifold and $X$ is a three dimensional submanifold of $Z$. Let $y = \gamma_x(t)$. 
A point $(y,t)$ lies in $Z^o$ if $t \in I^o$ and $x \in M^o(t)$. If $I = [t_0, t_1]$, the boundary of $Z$ is 
given by $\partial Z = X \cup (M(t_0), t_0) \cup (M(t_1), t_1)$. Define the projection 
$\mu: \mathbb{R}^3 \times I \to \mathbb{R}^3$ as $\mu(y,t) = y$. 
For $z \in Z$ and a point $w \in \mu(z)$, if $\mu^{-1}(w) \cap Z^o \neq \emptyset$ then $w \notin {\cal E}$. 
Hence a necessary condition for $w$ to be in ${\cal E}$ is that the line $\mu^{-1}(w)$ should be 
tangent to $\partial Z$.  For $x \in \cap_{i=1}^m F_i$, the cone of outward normals is ${ N}_x = \{ \sum_{i = 1}^{m} \alpha_i \cdot N_i$ \}, 
where $\sum_{i=1}^{m} \alpha_i = 1$, $\alpha_i \geq 0$ and $N_i$ is the outward normal to face $F_i \subset \partial M$ for $i=1, \ldots, m$.  
For $t \in I^o$, the cone of outward normals to $\partial Z$ at the point 
$(y,t)$ is given by ${\cal O} :=\{ \sum_{i = 1}^m \alpha_i \cdot (A(t) \cdot N_i, -g(x, N_i, t)) \}$.  Further, 
for $t = t_0$, the cone of outward normals to $\partial Z$ at the point 
$(y,t)$ is given by ${\cal P} := \{ \sum_{i = 1}^m \delta_i \cdot (A(t) \cdot N_i, -g(x, N_i, t)) - \beta \cdot \hat{e}_4 \}$, 
where $\hat{e}_4 = (0, 0, 0, 1)$ and $\beta, \delta_i \in \mathbb{R}$,  $\beta, \delta_i \geq 0$ for $i = 1, \ldots, m$
and $ \sum_{i = 1}^m \delta_i + \beta = 1$. Similarly, for $t = t_1$, 
the cone of outward normals to $\partial Z$ at the point $(y,t)$ is given by 
${\cal Q} := \{\sum_{i = 1}^m \delta_i \cdot (A(t) \cdot N_i, -g(x, N_i, t)) + \beta \cdot \hat{e}_4 \}$. 
Consider now case (i).  For $t = t_0$, if the line $\mu^{-1}(y)$ is tangent to a point $(y, t_0) \in \partial Z$, 
then there exists an outward normal to $\partial Z$ in ${\cal P}$ which is orthogonal to $\hat{e}_4$, i.e., 
there exist $\alpha_i \in \mathbb{R}$, $\alpha_i \geq 0$, and $\beta \in \mathbb{R}$, $\beta \geq 0$ such that 
$ \sum_{i=1}^m -  \delta_i \cdot g(x, N_i, t_0) = \beta \geq 0$.  In other words, there exists $n \in {N}_x$ 
such that $g(x, n, t_0) \leq 0$.  The proofs for case (ii) and case (iii) are similar.
\hfill $\square$

\section{Proof of Proposition~\ref{lsiProp}}		\label{proof2Sec}
\begin{figure}
 \centering
\includegraphics[scale=0.45]{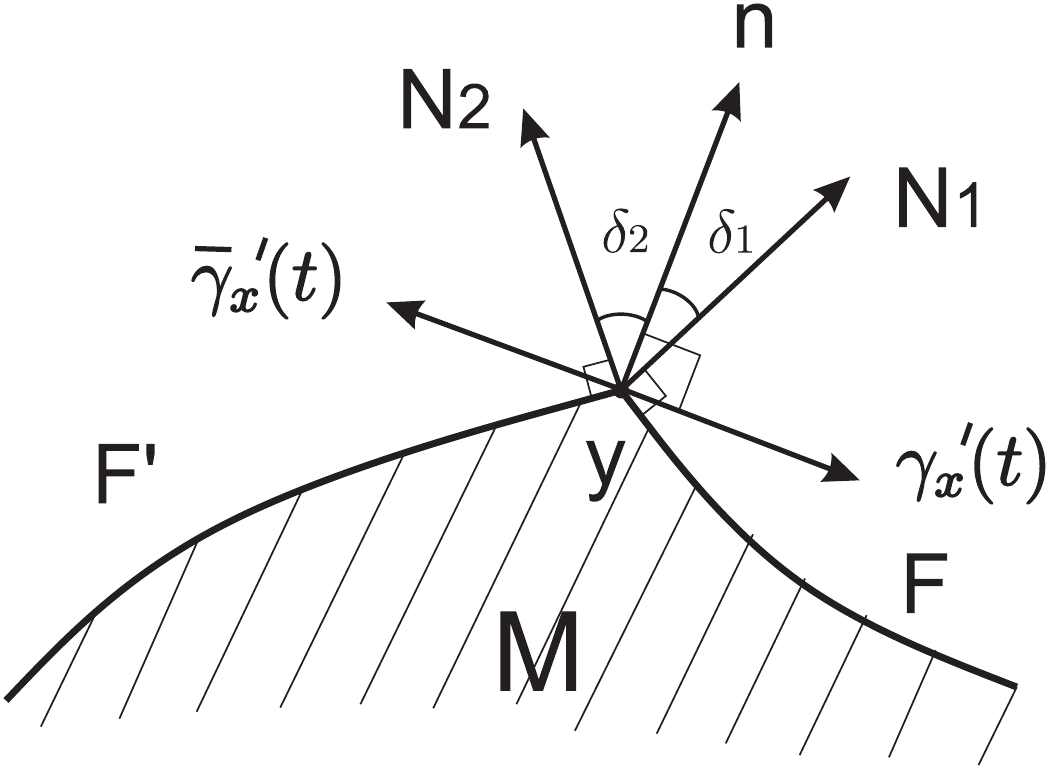}
 \caption{The inverse trajectory is in the exterior of $M$.}
 \label{lsiProofFig}
\end{figure}
\noindent {\em Proof.} 
Let $N_x$ be the cone of unit normals at $x \in E$ formed by $N_1$ and $N_2$, where $N_1$ and $N_2$  are the unique unit outward normals 
at $x$ to faces $F$ and $F'$ respectively.  Let $n \in N_x$ such that  $\left < A(t') \cdot n , \gamma_x'(t') \right > = 0$.  
Assume without loss of generality that $A(t') = I$ and $b(t') = 0$. 
Since $y$ is in the interior of face $C^E$, $n \neq N_1$ and $n \neq N_2$.  Suppose 
$n$ makes angles $\delta_1 > 0$ and $\delta_2 > 0$ with $N_1$ and $N_2$ respectively.  
Since $\gamma_x'(t') \bot n$, $\gamma_x'(t')$ makes angles $\delta_1$ and $\pi - \delta_2$ with faces $F$ and $F'$ respectively. 
It is easily verified that
 $\gamma_x(t') = \bar{\gamma}_x(t')$ and $\gamma_x'(t') = -\bar{\gamma}_x'(t')$, 
where $\bar{\gamma}_x'(t')$ is the derivative of the inverse trajectory of $x$.  
Hence $\bar{\gamma}_x'(t)$ makes angle $\delta_2$ with $F'$ and $\pi - \delta_1$ with $F$.
This is illustrated schematically in Fig.~\ref{lsiProofFig}.
The first order Taylor expansion of $\bar{\gamma}_x$ around $t'$ is given by 
$\bar{\gamma}_x(t' + \delta t) =  \bar{\gamma}_x(t') + \delta t \cdot \bar{\gamma}_x'(t')$.
Since $\bar{\gamma}_x'(t')$ points in exterior of solid $M(t')$, 
we conclude that for $\delta t$ small enough, the inverse trajectory $\bar{\gamma}_x(t)$ is in the exterior of solid $M(t')$ 
for all $t \in (t' - \delta t, t' + \delta t)$.  
\hfill $\square$

\end{document}